\documentstyle[elsart12,osa,epsf,eqsecnum,epsfig,graphics,cite]{revtex}
\begin{document}


\voffset1.5cm


\title{Particle Correlations in Saturated QCD Matter}
\author{Rudolf Baier $^{1}$, Alex Kovner$^{2,3}$, Marzia Nardi$^{2,4,5}$ and  Urs Achim Wiedemann$^{1,2}$}

\address{
$^1$Physics Department, University of Bielefeld, D-33501 Bielefeld, Germany\\[0pt]
$^2$ Physics Department, Theory Division, CERN, CH-1211 Geneva 23, Switzerland\\[0pt]
$^3$ Physics Department, University of Connecticut, 2152 Hillside
Road, Storrs, CT 06269-3046, USA\\[0pt]
$^4$Centro Studi e Ricerche "E.Fermi" , Compendio Viminale, I-00184 Rome, Italy\\[0pt]
$^5$INFN Sez. Torino, Via P. Giuria 1, I-10125 Torino, Italy}
\date{\today}
\maketitle

\begin{abstract}
We study quantitatively angular correlations in the two-particle spectrum produced by 
an energetic  probe scattering off a dense hadronic target with sizeable saturation 
momentum. To this end, two parton inclusive cross sections for arbitrary projectiles 
with small color charge density are derived in the eikonal formalism. Our results are the following:
For large momenta of the observed particles, the perturbative limit with characteristic
back-to-back correlation is recovered. As the trigger momenta get closer to 
the saturation scale $Q_s$, the angular distribution broadens. When the momenta are
significantly smaller than $Q_s$, the azimuthal distribution is broad but still peaked 
back-to-back. However, in a narrow momentum range $(0.5 \div 1.5)\, Q_s$, we 
observe that the azimuthal correlation splits into a double peak with maxima displaced
away from $180^\circ$. We argue that it is 
the soft multiple scattering physics that is responsible for the appearance of this shift 
in the angle of maximal correlation. We also point out that when the physical 
size of the projectile is particularly small, the double peak structure 
persists in a significantly wider range of final state momenta.
\end{abstract}

\section{Introduction}

High-$p_T$ near-side and back-to-back particle correlations have
become recently the focus of intensive study at the Relativistic
Heavy Ion Collider (RHIC). Such correlations
 give access to the microscopic dynamics of two major phenomena
searched for at RHIC in order to elucidate the properties of dense
QCD matter~\cite{Arsene:2004fa,Back:2004je,Adcox:2004mh,Adams:2005dq}: 
the phenomenon of perturbative saturation ~\cite{Mueller:1999yb,Blaizot:2004px,McLerran:2005kk,Iancu:2003xm,Weigert:2005us,Jalilian-Marian:2005jf}
expected to determine the {\it initial} condition of hadronic collisions at sufficiently high
center of mass energy, and the phenomenon of jet 
quenching~\cite{Jacobs:2004qv,Gyulassy:2003mc,Kovner:2003zj,Baier:2000mf} established
to suppress the high-$p_T$ hadronic yields due to {\it final} state interactions.
To disentangle initial and final state effects, it is clearly important to compare
data on nucleus-nucleus collisions to data on proton (deuteron)-nucleus
collisions in which final state effects are absent.

Two-particle correlations may provide further tests of the dynamics underlying
final state jet quenching, since it has been argued that their angular dependence 
is sensitive to the density of the produced matter and its collective 
flow~\cite{Salgado:2003rv,Armesto:2004pt}. Moreover,
it has been pointed out that in an ideal liquid, the energy radiated  off a hard final
state parton would propagate in a `sonic boom'~\cite{Casalderrey-Solana:2004qm}. 
This would lead to an azimuthal back-to-back correlation which is peaked away 
from $180^\circ$, and may thus serve as a characteristic signature of a medium 
with (almost) vanishing viscosity~\cite{Shuryak:2004cy}.
On the other hand, it has been suggested that initial state saturation could affect
dramatically the two-gluon spectrum~\cite{Kharzeev:2004bw} by strongly
reducing the angular correlation at high transverse momentum, leading practically
to a disappearance of the back-to-back correlations of jets in nuclear collisions. 
Although the effect is suggested to be
more dramatic for large nuclear projectiles, the trend should
also be seen for proton (deuteron)-nucleus collisions. 

The main purpose of the present work is to
study quantitatively saturation effects in the simplest process which
gives rise to two-parton correlation functions. In the target rest frame, this
is the emission of one gluon from a projectile quark $qA\rightarrow qgX$ 
with the transverse momentum of quark and gluon resolved.
These results are presented in Section~\ref{sec2}.

The other goal of this work is to set up a formalism for calculating differential 
multi-parton cross sections in the eikonal approximation~\cite{Kovner:2001vi}. In particular
we shall show how to take properly account of the final state radiation by 
unitarily transforming gluonic (and in general partonic) observables with the operator which 
dresses the partons by the cloud of Weizs\"acker-Williams gluons. We show that 
perturbative expansion of this cloud operator precisely generates the perturbative radiation in the final state.

The paper is organized as follows: In Section~\ref{sec2}, we discuss the physics of
two-parton correlation functions, starting from basic formulae and numerical results.
The subsequent sections discuss the general framework 
for calculating parton correlations in the eikonal formalism and some applications.  Section~\ref{sec3} presents the general discussion.
This formalism is then
applied to parton correlations for $q\, A \to q({\bf k})\, g({\bf p})\, X$ in Section~\ref{sec4}, 
and to $q\, A \to q\, g({\bf p}_1)\, g({\bf p}_2)\, X$ in Section~\ref{sec5}. The main
results are summarized in the Conclusions.

\section{Angular correlations in $q\, A \to q({\bf k})\, g({\bf p})\, X$}
\label{sec2}
We consider the simplest case of a two-particle final state.  
The projectile consists of a single quark. It scatters on a hadronic target
of sizeable saturation momentum and produces a  gluon in the final state: 
$q\, A \to q({\bf k})\, g({\bf p})\, X$. For a related discussion with strong
emphasis on breaking of $k_T$-factorization, see 
Refs.~\cite{Jalilian-Marian:2004da,Nikolaev:2005dd,Nikolaev:2004cu,Fujii:2005vj}.

\subsection{Basic formulae and definitions}
\label{sec2a}
The starting point of our study is the following expression for the doubly inclusive spectrum 
$q\, A \to q({\bf k})\, g({\bf p})\, X$ (for derivation see Sections~\ref{sec3} and~\ref{sec4}) 
\begin{eqnarray}
  {dN\over dy\, d{\bf k}\, d{\bf p}}
   &=& \frac{\alpha_s\, C_F}{\pi^2} \frac{1}{(2\pi)^4} 
   \int d{\bf z}\, d\bar{\bf z}\, d{\bf x}\, 
   e^{-i{\bf k}\cdot {\bf x}-i{\bf p}\cdot ({\bf z}-{\bar{\bf z}})}
   \frac{ ({\bf x} - {\bf z})\cdot \bar{\bf z}}
        { ({\bf x} - {\bf z})^2\,  {\bar{\bf z}}^2}
   \nonumber \\
   && \qquad  \times \left[ Q({\bf z},{\bf x},{\bf 0},\bar{\bf z}) 
                    S(\bar{\bf z},{\bf z}) + S({\bf x},{\bf 0})
                    \right.
   \nonumber \\
   && \qquad \left. \quad 
                    - S({\bf x},\bar{\bf z}) S(\bar{\bf z},{\bf 0})
                    - S({\bf x},{\bf z}) S({\bf z},{\bf 0}) \right]\, .
   \label{5.9x}
\end{eqnarray}
The entire information about the target nucleus is contained in two target 
averages of products of Wilson lines,
\begin{eqnarray}
  S({\bar {\bf u}},{\bf u}) &=& \langle \frac{1}{N} {\rm Tr}
  \left[ {W^F}^\dagger({\bar {\bf u}})  W^F({\bf u}) \right] \rangle_T\, , 
  \label{3.25}\\
  Q({\bar {\bf u}},{\bf u},{\bf z},{\bar {\bf z}}) 
  &=& \langle \frac{1}{N} {\rm Tr}
  \left[ {W^F}^\dagger({\bar {\bf u}})  W^F({\bf u}) 
         {W^F}^\dagger({\bf z}) W^F({\bar {\bf z}})  \right] \rangle_T \, .
  \label{3.26}
\end{eqnarray}
To model these target averages, we use the expressions~\cite{Mueller:1999yb,Jalilian-Marian:2004da}
\begin{eqnarray}
  S({\bar {\bf u}},{\bf u}) &=& \exp \left[ -v({\bar {\bf u}}-{\bf u})\right]\, ,
  \label{5.1}\\
  Q({\bar {\bf y}},{\bf x},{\bar {\bf x}},{\bf y}) 
    &=& 
    \frac{v({\bf x}-{\bar {\bf x}}) + v({\bf y}-{\bar {\bf y}})
              - v({\bf x}-{\bf y}) 
                 - v({\bar {\bf x}}-{\bar {\bf y}})   }
          {v({\bf x}-{\bar {\bf x}}) + v({\bf y}-{\bar {\bf y}})
               - v({\bf x}-{\bar {\bf y}}) - v({\bf y}-{\bar {\bf x}}) }
    \nonumber \\
    && \qquad \times
          \exp\left[ -v({\bf x}-{\bar {\bf x}}) 
                 - v({\bf y}-{\bar {\bf y}}) \right]
    \nonumber \\
    &&  - \frac{v({\bf x}-{\bar {\bf y}}) + v({\bf y}-{\bar {\bf x}})
              - v({\bf x}-{\bf y}) 
                 - v({\bar {\bf x}}-{\bar {\bf y}})   }
          {v({\bf x}-{\bar {\bf x}}) + v({\bf y}-{\bar {\bf y}})
               - v({\bf x}-{\bar {\bf y}}) - v({\bf y}-{\bar {\bf x}}) }
    \nonumber \\
    &&  \qquad \times
          \exp\left[ -v({\bf x}-{\bar {\bf y}}) 
                 - v({\bf y}-{\bar {\bf x}}) \right]\, ,
              \label{5.7}\\
  v({\bf x}) &=& {\bf x}^2\, 
      {\tilde{Q}_{s}^2({\bf x})\over 8} \equiv {\bf x}^2\, 
      {Q_{s,0}^2\over 8} 
     \log\left[ \frac{1}{{\bf x}^2 \Lambda^2} + a \right]\, .
     \label{5.2}
\end{eqnarray}
The function $v({\bf x})$ has the meaning of the target gluon field correlation function and 
is directly proportional to the gluon density in the target, hence the logarithmic dependence 
on the transverse separation.

For numerical evaluation, we shall take $\Lambda \equiv \Lambda_{\rm QCD} = 
0.2\, {\rm GeV}$. The small regulator $a=1/(x_c^2\, \Lambda^2)$, 
$x_c = 3\, {\rm GeV}^{-1}$ is chosen such that the logarithm in (\ref{5.2}) 
does not turn negative and that the sensitivity on the infrared cut-off 
is negligible for sufficiently large momentum~\cite{Baier:2003hr}. The saturation scale 
$Q_s$ is defined implicitly as
\begin{equation}
        Q_s^2 \equiv \tilde{Q}_s^2({\bf x}^2 = 1/Q_s^2)\, .
        \label{5.3}
\end{equation}
With this definition, $Q_s^2 = 2\, {\rm GeV}^{2}$ corresponds to
$Q_{s,0}^2 \simeq 0.5\, {\rm GeV}^{2}$ in Eq. (\ref{5.2}).

In the model (\ref{5.1})-(\ref{5.2})
the properties of the single inclusive gluon radiation spectrum $q\, A \to q\, g({\bf p})\,  X$ 
and its small-$x$ evolution have been studied 
extensively in recent years and are well understood~\cite{Kharzeev:2002pc,Baier:2003hr,Jalilian-Marian:2003mf,Kharzeev:2003wz,Albacete:2003iq}. This spectrum is obtained from the two-particle correlation 
function (\ref{5.9x}), integrating over the quark momentum ${\bf k}$,
\begin{eqnarray}
  {dN\over dy\, d{\bf p}}
   &=& \frac{\alpha_s\, C_F}{\pi^2} \frac{1}{(2\pi)^2} 
   \int d{\bf z}\, d\bar{\bf z}\,
   e^{-i{\bf p}\cdot ({\bf z}-{\bar{\bf z}})}
   \frac{ {\bf z}\cdot \bar{\bf z}}
        { {\bf z}^2\,  {\bar{\bf z}}^2}
   \nonumber \\
   && \qquad  \times \left[
                    S^2(\bar{\bf z},{\bf z}) + 1
                                        - S^2(\bar{\bf z})
                    - S^2({\bf z}) \right]\, .
   \label{5.9xx}
\end{eqnarray}
In the approximation (\ref{5.1})-(\ref{5.2}), one finds
\begin{eqnarray}
 \frac{dN}{dy\, d{\bf p}}
  &=& \frac{1}{\pi} \, \int  \, d{\bf z}\,
      d\bar{\bf z}  \frac{1}{(2 \pi)^2}  \frac{\alpha_s C_F}{\pi}
      \frac{{\bf z} \cdot \bar{\bf z}}{{\bf z}^2 {\bar{\bf z}}^2} \,
      e^{-i {\bf p}\cdot ({\bf z}-\bar{\bf z})} \,
  \nonumber \\
   && \times
   \left[ 1+e^{ - 2v({\bf z}-\bar{\bf z}) }  -  e^{- 2v({\bf z}) } - 
               e^{- 2v(\bar{\bf z}) }  \right]\, ,
    \label{5.4}
\end{eqnarray}
which coincides with the quasiclassical expression of \cite{Kovchegov:1998bi}.

In the limit of small transverse momenta, the relevant values of the 
transverse coordinates ${\bf z}$, $\bar {\bf z}$ are large. In this case, 
the logarithm in (\ref{5.2}) becomes unimportant and the Gaussian
approximation $Q_s^2({\bf x}) = \overline{Q}_s^2$ is justified.
Formally, one finds in this approximation
\begin{equation}
   \frac{dN}{dy\, d{\bf p} }
   \Bigg\vert_{\rm Gaussian} = 
     \frac{\alpha_s\, C_F}{\pi^3 \overline{Q}_s^2}
     \int d{\bf q}\, 
        e^{- {\bf q}^2/\overline{Q}_s^2}\, 
        \frac{{\bf q}^2}{{\bf p}^2\, \left({\bf p}-{\bf q} \right)^2}   \, .
       \label{5.6}
\end{equation}
Here, $\frac{{\bf q}^2}{{\bf p}^2\, \left({\bf p}-{\bf q} \right)^2}$ is the typical momentum 
dependence of gluons radiated off a high-energy quark, which received a momentum 
transfer ${\bf q}$. For gluons of small transverse 
momentum ${\bf p}$, the momentum transfer ${\bf q}$ from the target is 
accumulated according to transverse Brownian motion 
$\langle {\bf q}^2 \rangle \propto \overline{Q}_s^2 \propto A^{1/3}$. 
For a hard gluon $\vert {\bf p}\vert \gg Q_s$, however,  Eq. (\ref{5.4}) shows
the typical power law $\propto \frac{1}{{\bf p}^4} \ln \frac{{\bf p}^2}{\Lambda^2}$
characteristic for a single hard scattering off the power-law tail of a hard scattering
center, see (\ref{5.5}) below. In this 
way, the ansatz (\ref{5.2}) for the saturation momentum characterizes a medium 
with a physically sensible 
interpolation between single hard and multiple soft scattering as a function 
of transverse momentum transfer. We note that this is the QCD analogue of
QED Moli\`ere scattering theory~\cite{Bethe}, which achieves the analogous interpolation between
soft multiple and single hard target momentum transfers for electromagnetic
scattering processes.

\subsection{Perturbative baseline}
As a baseline for comparison of our numerical results, we collect here the perturbative 
expressions for the single and double inclusive cross sections. These do not include 
the effects of multiple rescattering which are due to the existence of a large saturation 
scale in the target. The perturbative baseline is obtained by identifying the single hard 
interaction term in the target averages (\ref{5.1}), (\ref{5.7}). This corresponds to an
expansion to first order in $v({\bf x})$:
\begin{eqnarray}
  S({\bar {\bf u}},{\bf u}) &=& 1-v({\bar {\bf u}}-{\bf u}) + O(v^2)\, ,
  \label{5.4x}\\
  Q({\bar {\bf u}},{\bf u},{\bar {\bf w}},{\bf w})
  &=& 1 - \left[ v({\bar {\bf u}}-{\bf u}) + v({\bar {\bf w}}-{\bf w})
                   + v({\bar {\bf u}}-{\bf w}) + v({\bar {\bf w}}-{\bf u})
                   \right.
  \nonumber \\
  && \qquad \left. 
     -v({\bf u}-{\bf w}) - v({\bar {\bf u}}-{\bar{\bf w}}) \right] + O(v^2)\, .
  \label{5.5x}
\end{eqnarray}
Inserting this into Eq.(\ref{5.9x}), we find
\begin{equation}
   {dN\over dy\, d{\bf k}\, d{\bf p}}
   =\frac{\alpha_s\, C_F}{\pi^2}  \frac{Q^2_{s,0}}{\pi}
            \frac{1}{{\bf p}^2 ({\bf k}+{\bf p})^2{\bf k}^2}\, .
   \label{pert1}
\end{equation}
To obtain (\ref{pert1}), one has to keep the logarithmic
dependence in the scale entering $v({\bf x})$ in (\ref{5.2}). 
Neglecting this dependence of $Q_s$ on ${\bf x}$
amounts to neglecting the possibly large (albeit more rare) momentum transfer due to the 
interaction with the high momentum tail of the target fields. In this case, 
$Q_s^2({\bf x}) = \overline{Q}_s^2$, one finds instead
\begin{equation}
   {dN_{\rm Gauss}\over dy\, d{\bf k}\, d{\bf p}}
   =\frac{\alpha_s\, C_F}{\pi^2}  \frac{\overline{Q}^2_{s}}{{\bf p}^4}
        \delta^{(2)}\left({\bf p}+{\bf k}\right)\, ,
   \label{pert1gauss}
\end{equation}
where the quark and gluon momenta are exactly balanced. This shows
that as long as the interaction with the target can not impart a significant 
kick to a propagating parton, large momenta in the final state can only appear 
due to the large relative momentum between the quark and the gluon in the initial wave 
function of the projectile. 

Integrating Eq.(\ref{pert1}) over the quark mometum ${\bf k}$, we obtain the perturbative 
expression for the single gluon inclusive cross section
\begin{eqnarray}
   \frac{dN}{dy\, d{\bf p} }
   \Bigg\vert_{{\vert{\bf p}\vert} \gg Q_s}
  = 
  \frac{2 \alpha_s\, C_F\, Q_{s,0}^2}{\pi^2}
  \frac{1}{{\bf p}^4} \, (\ln [\frac{{\bf p}^2}{4 \Lambda^2}]
  +2 \gamma_E - 1)\, .
  \label{5.5}
\end{eqnarray}
To determine the distribution of the total recoil momentum, we integrate
(\ref{5.9x}) over the relative momentum of the quark-gluon pair in the final state. 
Defining 
\begin{equation}
{\bf K} = {\bf k} + {\bf p} \, , \qquad  {\bf q} = {\bf k}-{\bf p}\, ,
\end{equation}
we find to leading logarithmic accuracy
 \begin{equation}
 {dN\over dy\, d{\bf K}} = 2 \frac{\alpha_s\, C_F}{\pi^2}\, 
                              \frac{Q^2_{s,0}}{({\bf K}^2)^2}
                             \ln \left[ {{\bf K}^2\over \Lambda^2} \right]\, .
 \label{pert2}
\end{equation}
This expression exhibits the typical perturbative power-law dependence for 
the total recoil momentum $K$ above the infrared regulator scale $\Lambda$. 
This is consistent with the expectation, that for large recoil ${\bf K}$, the target
behaves like a single hard perturbative scattering center. 

We also give here the perturbative expression for the two 
gluon inclusive cross section $q\, A \to q\, g({\bf p}_1)\, g({\bf p}_2)\, X$,
with gluons produced at rapidities $\eta$ and $\xi$. The full expression, including the effects of the saturation is given in Eq.(\ref{4.13}) 
below. Its perturbative limit is
\begin{equation}
  {dN\over d\eta d{\bf p}_1d\xi d{\bf p}_2} 
  = {\alpha_s^2\over \pi^6} Q^2_{s,0}{1\over {\bf p}_1^2{\bf p}_2^2({\bf p}_1+{\bf p}_2)^2}
  \ln{({\bf p}_1+{\bf p}_2)^2\over \Lambda^2}\, .
  \label{5.8x}
\end{equation}
%
\subsection{Distribution of the recoil momentum}
The total transverse momentum ${\bf K}$ of the quark-gluon pair in 
the final state traces the distribution of recoil momentum transferred from the target.
We calculate it by integrating (\ref{5.9x}) over the relative transverse momentum
${\bf q} $. After performing some angular integrations, one finds
\begin{eqnarray}
{dN\over dy\, K\, dK}
   &=& \frac{\alpha_s\, C_F}{\pi^2} 
   \int_0^\infty dz\, \int_0^{1/\Lambda_{\rm cut}} d\bar z\,
   \int_0^{2\pi} d\phi\,  \frac{z}{ \bar z }\, J_0(Kz) 
   \nonumber \\
   && \qquad  \times \left[ Q({\bf z}+\bar{\bf z},{\bf z},{\bf 0},\bar{\bf z}) 
                    S({\bf z}) + S({\bf z})
                    \right.
   \nonumber \\
   && \qquad \left. \quad 
                    - S({\bf z}-\bar{\bf z}) S(\bar{\bf z})
                    - S(\bar{\bf z}) S({\bf z}+\bar{\bf z}) \right]\, .
   \label{5.12x}
\end{eqnarray}
Here, we have introduced the infrared cut-off $\Lambda_{\rm cut}$ to regulate the logarithmically 
infrared divergent $\bar{z}$-integral. For the discussion in this subsection, we choose
$\Lambda_{\rm cut} = \Lambda$, the same infrared cut-off, which regulates 
the perturbative expression (\ref{pert2}). However, as discussed below, $\Lambda_{\rm cut}$
has a physical interpretation as regulator of the transverse size of the incoming projectile.
We will discuss the implications of restricing the size of the projectile in more detail later. 

\begin{figure}[h]\epsfxsize=10.7cm
\centerline{\epsfbox{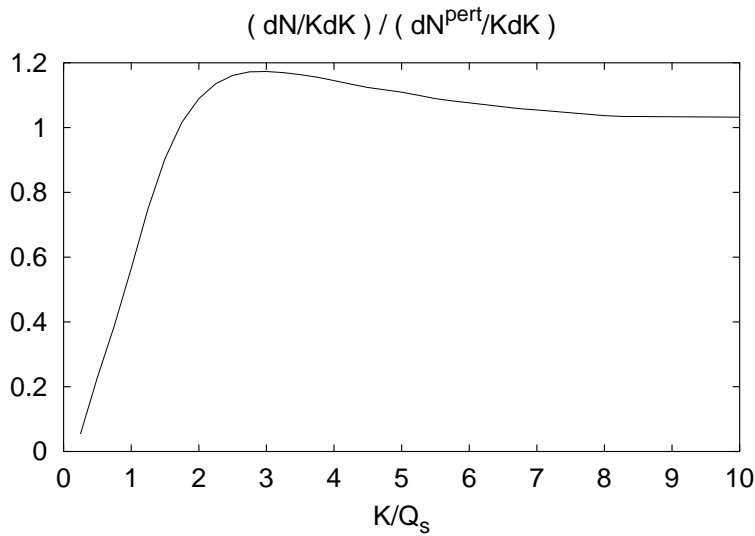}}
\vspace{0.5cm}
\caption{ The full two-parton (quark-gluon) correlator (\protect\ref{5.12x}) 
normalized to its perturbative limit (\protect\ref{pert2}) as a function of the total pair
momentum $K = \vert {\bf k} + {\bf p}\vert$. 
}\label{fig1}
\end{figure}

We have evaluated expression (\ref{5.12x}) numerically. In Fig.~\ref{fig1} we plot the ratio of the distribution (\ref{5.12x}) to the perturbative expression
 (\ref{pert2}). For momenta $K\gg Q_s$ well above the saturation scale, the recoil distribution 
 (\ref{5.12x}) approaches the perturbative power law $\propto \frac{1}{K^4} \ln \frac{K^2}{\Lambda^2}$
and is indistinguishable from the perturbative expression. For low momenta $K< Q_s$, the saturated 
distribution is suppressed relative to the perturbative one. This is the 
manifestation of the fact that a parton propagating through the saturated target is very unlikely to 
get a kick of momentum less than of order of $Q_s$,
whereas in the interaction with the perturbative target on the contrary low momentum transfer processes 
are very important. Finally, Fig.~\ref{fig1} exhibits a fairly wide maximum 
around $K=(2\div 3)\,Q_s$, which is due to the transfer of momentum of order $Q_s$ to one of the propagating 
partons ($q$ or $g$). The general features of Fig.~\ref{fig1} are very similar
to those of the nuclear modification factor calculated within the same model 
in~\cite{Baier:2003hr,Kharzeev:2003wz,Jalilian-Marian:2003mf}.

%
\subsection{Angular Correlations - large trigger momentum}
We next study the angular dependence of the two-parton correlation function
(\ref{5.9x}). Shifting ${\bf z} \to {\bf z} + {\bf x}$ and introducing radial coordinates, 
we find
\begin{eqnarray}
  {dN\over dy\, kdk\, pdp\, d\Delta\phi_{kp}}
   &=& \frac{\alpha_s\, C_F}{\pi^2} \frac{1}{(2\pi)^2} 
   \int x\ dx\, \int^{\frac{1}{\Lambda_{\rm cut}}} dz\, \, \int^{\frac{1}{\Lambda_{\rm cut}}} 
   d\bar z\, 
   \nonumber \\
  && \times  \int d\phi_z\, d\phi_{\bar z}\, 
    \cos(\phi_z -\phi_{\bar z})\, J_0\left( \sqrt{A^2 + B^2}\right)
   \nonumber \\
   && \qquad  \times \left[ Q({\bf z}+{\bf x},{\bf x},{\bf 0},\bar{\bf z}) 
                    S(\bar{\bf z},{\bf z}+{\bf x}) + S({\bf x},{\bf 0})
                    \right.
   \nonumber \\
   && \qquad \left. \quad 
                  - S({\bf x},\bar{\bf z}) S(\bar{\bf z},{\bf 0})
                  - S({\bf 0},{\bf z}) S({\bf z}+{\bf x},{\bf 0}) 
                  \right]\, .
   \label{5.17}
\end{eqnarray}
Here, ${\bf z}\cdot \bar{\bf z} =  z\bar z \cos(\phi_z -\phi_{\bar z})$,
${\bf x}\cdot {\bf z} = x z \cos(\phi_z )$, $\bar{\bf z}\cdot {\bf x} = \bar zx \cos(\phi_{\bar z} )$
and the notational shorthands
\begin{eqnarray}
  && A = kx\cos(\Delta \phi_{kp}) + px + pz\cos(\phi_z) - p\bar z\cos(\phi_{\bar z})\, ,
  \label{5.18} \\
  && B = kx\sin(\Delta \phi_{kp}) - pz\sin(\phi_z) + p\bar z\sin(\phi_{\bar z})\, .
  \label{5.19}
\end{eqnarray}
The $z$ and $\bar z$ integrals in equation (\ref{5.17}) are logarithmically IR divergent. 
Regulating these integrals by cutting off both integrations at distances greater than 
$1/\Lambda_{\rm cut}$, we can numerically determine the correlation of the two partons 
as a function of their relative azimuthal angle $\Delta\phi_{kp}$ and their transverse momentum
$k$ and $p$. Technically,  it is sufficient to regulate either one of the integrations $z$ or $\bar z$, 
to arrive at a finite result. Physically, the cut-off $1/\Lambda_{\rm cut}$ limits the transverse
size of the incoming wave function to $z < 1/\Lambda_{\rm cut}$ in the amplitude and to 
$\bar{z} < 1/\Lambda_{\rm cut}$ in the complex conjugate amplitude. Thus, we choose
to implement the cut-off in a symmetric way.

\begin{figure}[h]
\begin{center}
\includegraphics[width=14.5cm]{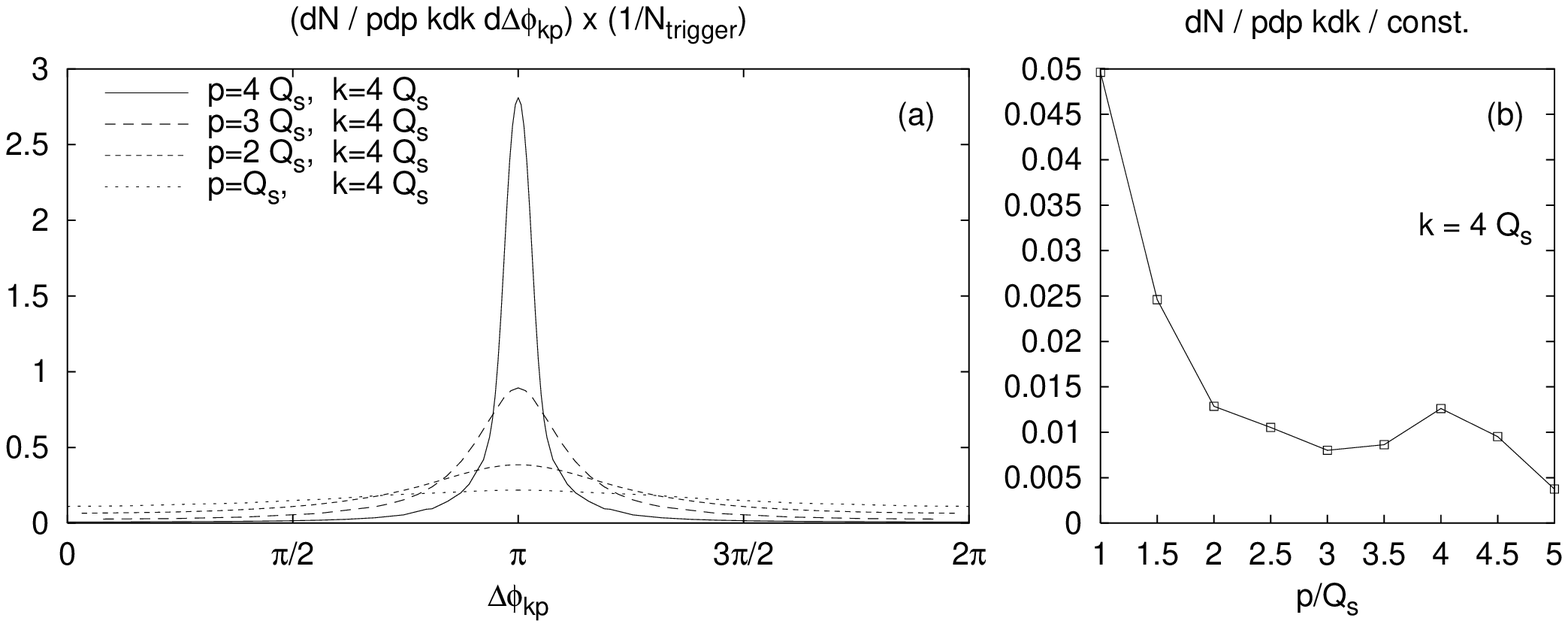}
\end{center}
\vspace{-1.0cm}
\begin{center}
\includegraphics[width=14.5cm]{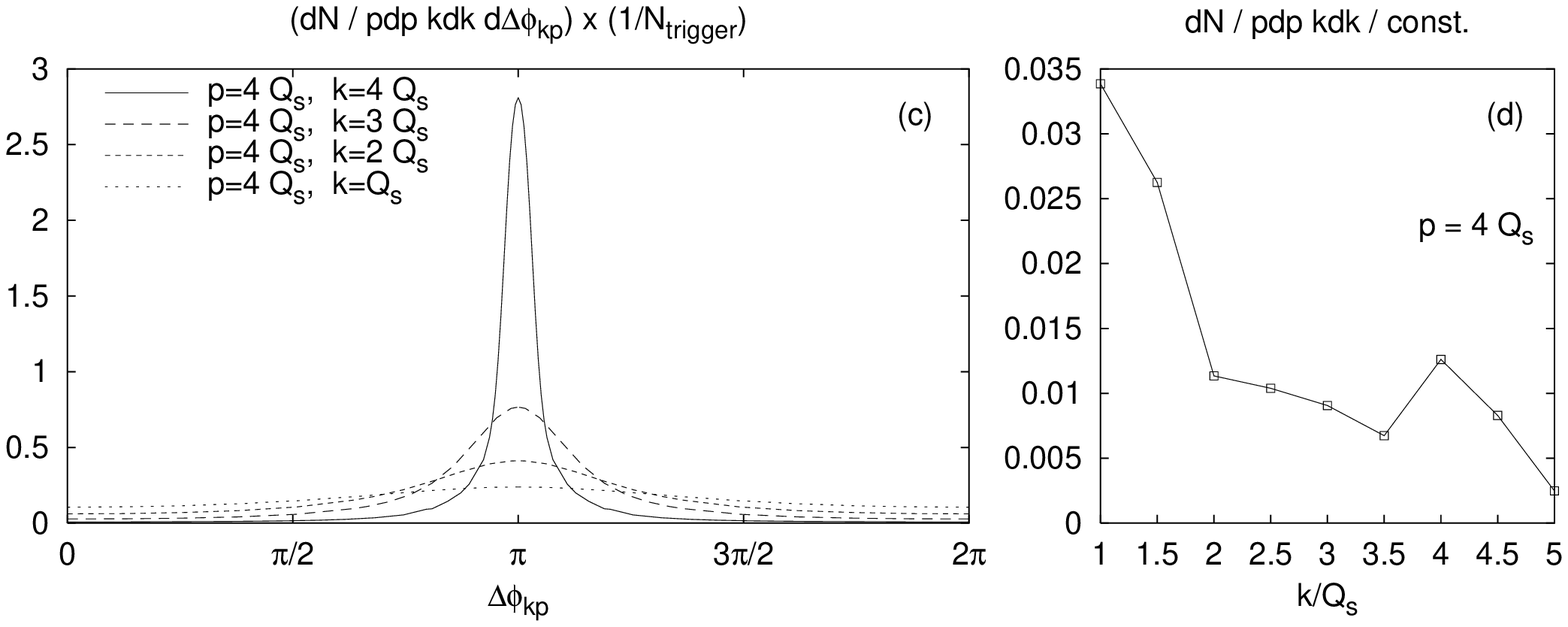}
\end{center}
\caption{(a),(c): The azimuthal dependence of the quark-gluon correlator (\ref{5.17})
for different values of quark ($k$) and gluon ($p$) transverse momentum. The area
under the curves is normalized to one, as required for differential rates per trigger
particle. (b),(d): The value of the $\Delta \Phi_{kp}$-integrated double differential
two-parton yield (\ref{5.17}), ${\rm const.} = \frac{\alpha_s\, C_F}{\pi^2} \frac{1}{(2\pi)^2}$.
}\label{fig2}
\end{figure}

Fig.~\ref{fig2}(a) shows the correlation function as a function of the azimuthal angle for a fixed 
and relatively large value of the quark momentum $k=4Q_s$ and values of the gluon 
momentum $p$ between $Q_s$ and $4Q_s$. We observe that for larger values of $p$ the 
distribution exhibits a strong back-to-back correlation. As $p$ decreases, the distribution 
becomes wider, and finally at $p=Q_s$ it is practically flat.

The main features of Fig.~\ref{fig2}(a) can be understood in the following simple picture. The 
incoming wave function of the projectile in the present approximation contains only one 
quark and up to one gluon (see Sections~\ref{sec3}-\ref{sec5}). The total transverse 
momentum of the incoming quark-gluon system is zero, and thus they are exactly correlated 
back-to-back. While travelling through the target, each parton gets a transverse kick which 
in most cases is close to $Q_s$. However, albeit with small probability, the impinging partons 
can scatter off the perturbative high momentum tail of the target field (Moli\`ere scattering). 
The final state configurations with large transverse momentum of both, quark and gluon, come 
mainly from two sources: either from the wave function components of the projectile 
where $q$ and $g$ have both large momentum and experience soft momentum transfer, or 
from a low momentum component of the incoming wave function where the quark experiences 
hard scattering off the Moli\`ere tail and subsequently radiates a gluon in the final state.
[In the present calculation only the scattered quark can radiate. The radiation off the gluon 
is higher order correction in $\alpha_s$, since the weight of the gluon component in the initial 
state itself is of order $\alpha_s$.] These events are therefore mostly back-to-back 
correlated, mirroring correlations in the initial projectile wave function and the back-to-back 
nature of the final state radiation.

On the other hand, final states with a large transverse momentum of the quark but a 
gluon with $p=Q_s$ can only arise due to Moli\`ere scattering of the quark, since there are no 
asymmetric configurations in the initial projectile wave function. The direction of the hard kick 
the quark experiences is random and uncorrelated with the direction of momentum of the gluon, 
and thus the angular distribution in this regime is almost completely flat, see Fig.~\ref{fig2}(a). 
For the intermediate values of $p$, the 
distribution interpolates between these two extremes.

The angular integrated intensity (\ref{5.17}) is shown in Fig.~\ref{fig2}(b) as a function of $p$ 
for a trigger quark momentum $k=4\,Q_s$. It exhibits a characteristic maximum at $p=0$ and 
another less sharp maximum at $p=k$. The maximum at $p=0$ originates from the 
momentum distribution of the 
incoming projectile wave function, which is peaked at small values of momentum as $1/p^2$. 
The behavior near this peak is unaffected by the presence of $Q_s$, but it is affected by
the infrared cut-off $\Lambda_{\rm cut}$, which eliminates the low relative momentum configurations 
from the initial wave function. 
The maximum at $p=k$ can be traced to the pertubative expression (\ref{pert1}) which (after
integration over $\Delta \phi_{kp}$) diverges linearly at $p=k$. This divergence originates from 
the back-to-back correlations between the quark and the gluon in the projectile wave function 
and the back-to-back final state radiation. For the saturated target however, the final state 
momentum of each parton is smeared around its initial state value within an interval of the 
width $\pm Q_s$. Thus the linearly divergent peak of (\ref{pert1}) turns into a finite peak of 
width $Q_s$ in Fig.~\ref{fig2}(b).

Figs.~\ref{fig2}(c,d) present analogous plots, where the momentum of the 
outgoing gluon is fixed at $p=4Q_s$ and the quark momentum is varied. Although some fine 
details are different, the overall picture is qualitatively very similar to Figs.~\ref{fig2}(a,b).

\subsection{Angular Correlations - trigger momenta close to $Q_s$}
\begin{figure}[h]
\begin{center}
\includegraphics[width=8.5cm]{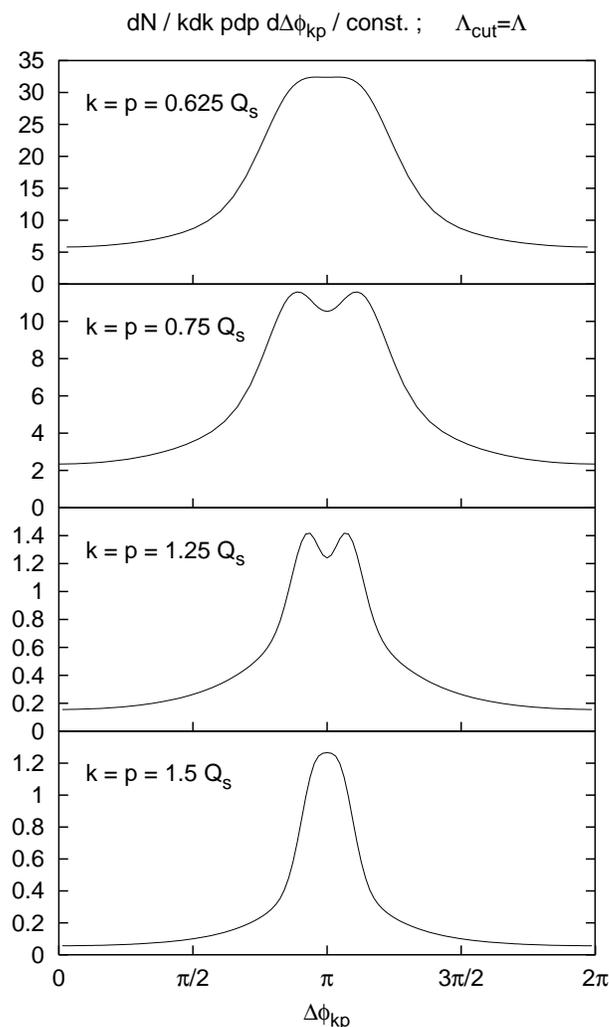}
\end{center}
\caption{Angular dependence of the quark-gluon correlator (\ref{5.17}) for different
values of the trigger momentum close to $Q_s$, ${\rm const.} = \frac{\alpha_s\, C_F}{\pi^2} \frac{1}{(2\pi)^2}$.
}\label{fig3}
\end{figure}

We consider now the regime where the trigger momenta are of the order of $Q_s$.
We concentrate on the situation when the associated momenta are equal, $p=k$.
In Fig.~\ref{fig3}, we show angular correlations when the trigger momenta are varied 
between $0.625\, Q_s$ and $1.5\, Q_s$. These plots show a structure distinct 
from the expected back-to-back correlations. The angular correlation is not peaked at 
$\phi=\pi$, but instead at $\phi=\pi-\delta$, where $\delta$ is clearly greater for smaller 
values of $p$. The dip at $\phi=\pi$ disappears for trigger momenta lower than about 
$0.6\, Q_s$ and higher than about $1.4\, Q_s$.

The dip at $\pi$ arises due to a coherent scattering effect, 
namely due to soft multiple scattering of the initial quark-gluon components 
of small transverse size. If the size 
of the pair is smaller than the transverse correlation length of the target fields, then
we expect that the pair scatters as one single object as it sees the same target fields. 
It then picks up a typical soft momentum which is shared equally between the 
quark and the gluon. [In the model defined by (\ref{5.2}), it seems reasonable to use
$Q_{s,0}$ as the typical soft scale, since the logarithmic correction which shifts 
$Q_{s,0}$ to $Q_s$ is due to harder target gluons. Also, the correlation length of 
the target may be estimated to be of order $1/Q_{s,0}$.]
In the initial state the quark and the gluon have momenta
equal in magnitude and opposite in direction. In our trigger momenta kinematics, where 
the magnitudes of the trigger momenta are also equal, the momentum 
transfer from the target must be in the direction perpendicular to the initial momenta. We 
can thus estimate the maximal correlation angle when $p>Q_{s,0}$ by the following
argument. We start with the initial momenta of the quark and the gluon ${\bf k}_{in}$, $-{\bf k}_{in}$.
 In the final state ${\bf k}={\bf k}_{in}+\delta {\bf k}$ and 
 ${\bf p}=-{\bf k}_{in}+\delta {\bf k}$ with $|\delta {\bf k}|={Q_{s,0}/2}$ and 
 ${\bf k}_{in}\cdot \delta{\bf k}=0$. 
For $k\gg Q_{s,0}$ we find the angle between the momenta in the final state 
\begin{equation}
  \phi=\pi-{Q_{s,0}\over {\sqrt 2}k}\, .
  \label{shift}
\end{equation}
The magnitude of the effect seen in Fig.~\ref{fig3} is consistent with this rough 
estimate. 
%
%
So far, we have discussed how coherent scattering of incoming parton pairs can lead to
a shift of the backward peak away of zero. To understand why the dip in the two-parton
correlator appears only for a narrow range of trigger momenta, we consider
two other scattering mechanisms: i) the perturbative hard Moli\`ere scattering 
off the large momentum tails of the target, which leads to a back-to-back 
distribution, and ii) the incoherent soft scattering for those components of the wave function,
whose transverse size is greater than the correlation length $\sim 1/Q_{s,0}$ of the target fields.
Indeed, when the trigger momenta are much higher than $Q_s$, the phase space in the initial state 
from which one can get to these final momenta by the coherent scattering mechanism is very small. 
Thus the perturbative final state radiation dominates, which is exactly back-to-back. The 
phase space for coherent scattering of incoming parton pairs also disappears if $k_{trigger}$ 
is significantly smaller than $Q_s$, since the coherent scattering mechanism requires enough 
particles in the initial state with momenta $k_{initial}<k_{trigger}-\beta Q_s$ with $\beta$ - a 
number of order one. Thus, it seems natural that the coherent scattering mechanism is dominant 
only for trigger momenta around $Q_s$.

To further test the mechanism suggested to underly the dip in the two-parton correlation
functions, one can exploit the dependence of the two-parton correlation function (\ref{5.17})
on the infrared cut-off $\Lambda_{\rm cut}$. In the derivation of this expression in
Section~\ref{sec4}, this cut-off dependence can be seen to restrict the cross section
of the final state radiation. Thus, if one artificially raises the infrared cut-off so that it is 
close to the trigger momentum, one expects that most of the contribution of the
back-to-back correlated, final state radiation is eliminated. One expects a dip at 
$\phi=\pi$ for a much larger trigger momentum, with maxima at a position given by 
the simple estimate (\ref{shift}). We have performed various such consistency checks to
further substantiate the explanation given above and find that the results conform with this simple picture.

\begin{figure}[h]
\begin{center}
\includegraphics[width=8.5cm]{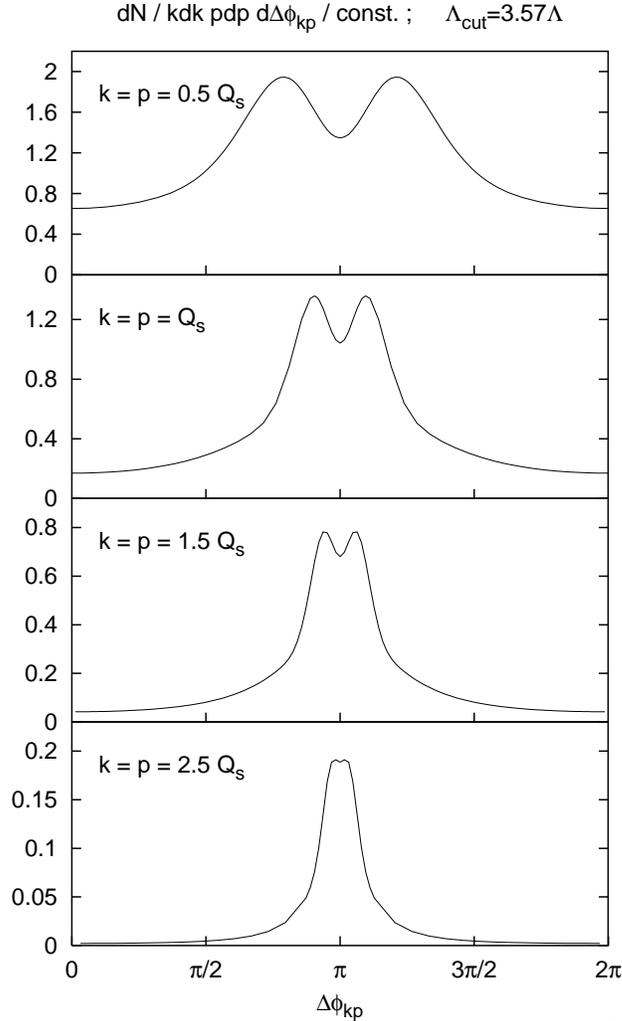}
\end{center}
\caption{Angular dependence of the quark-gluon correlator (\ref{5.17}) for different
values of the trigger momentum close to $Q_s$, ${\rm const.} = \frac{\alpha_s\, C_F}{\pi^2} \frac{1}{(2\pi)^2}$. Here, the transverse size of the incoming projectile wavefunction is
restricted to a small value $1/\Lambda_{\rm cut}$.
}\label{fig4}
\end{figure}

The physical interpretation to the cut-off dependence of (\ref{5.17}) suggests that one can use it to extract some additional information from our calculation.
As mentioned above, and as can be seen from the derivation in Section~\ref{sec4}, 
the cut-off on the $z$ and 
$\bar z$ integrals restricts the size of the projectile wave function to $1/\Lambda_{\rm cut}$.
Thus, varying the cut-off, we can probe the physics of small size projectiles. For example,
since the scale of $0.3$ Fermi is the natural size of the constituent quark, one may ask
what is the effect of restricting the gluon in the wave function to be emitted no further 
than $0.3$ Fermi from the valence quark. [Alternatively, one may think of this excercise 
as a ``poor man's'' way to mimick the doubly inclusive spectrum of a scattering of a 
dipole of this size.] In any case, raising the cut-off to this value should give a reasonable 
indication of the upper limit on the effect that one might expect.
The numerical results with the cut-off $\Lambda_{\rm cut}=Q_{s,0}=3.57\, \Lambda$ are 
shown in Fig.~\ref{fig4}. 
In accordance with our expectations, the effect is more pronounced. The dip at $\phi=\pi$ 
now persists up to higher values of the trigger momentum, about $2.5\, Q_s$. It is also much 
deeper than for the lower IR cut-off.
%
%
This concludes our discussion of the numerical results. We now turn to the derivation of the basic formulae used in this section and a more general discussion of differential cross sections in the eikonal formalism.
\section{Differential cross sections in the eikonal approximation}
\label{sec3}

Below we follow the basic formalism of \cite{Kovner:2001vi}. Consider an
energetic hadronic projectile impinging on a large nuclear target.
The projectile is characterized by a wave function, in which the
relevant degrees of freedom are the transverse positions and
color states of the partons,
\begin{equation}
  \vert \Psi_{in} \rangle = \sum_{\{\alpha_i,{\bf x}_i\}}\, 
  \psi(\{\alpha_i, {\bf x}_i\})\,
  \vert\{\alpha_i,{\bf x}_i\}\rangle\, .
  \label{2.1}
\end{equation}
The color index $\alpha_i$ can belong to the fundamental,
antifundamental or adjoint representation of the color $SU(N_c)$
group, corresponding to quark, antiquark or gluon in the
wavefunction. In what follows we will consider wave functions with
a small number of partons.

At high energy, the propagation time through the target is short,
and thus partons propagate independently of each other. For the
same reason the transverse positions of the partons do not change
during the propagation. The only effect of the propagation is that
the wave function of each parton acquires an eikonal phase due to
the interaction with the gluon field of the target. Thus the
projectile emerges form the interaction region with the wave
function
\begin{equation}
  \vert \Psi_{out} \rangle ={\cal S} \vert \Psi_{in} \rangle=
  \sum_{\{\alpha_i,{\bf x}_i\}}\psi(\{\alpha_i, {\bf x}_i\})
  \prod_iW({\bf x}_i)_{\alpha_i \beta_i}\, 
  \vert\{\beta_i,{\bf x}_i\}\rangle\, .
  \label{2.2}
\end{equation}
Here ${\cal S}$ is the $S$-matrix, and the $W$'s are Wilson lines
\begin{equation}
  W({\bf x}_i)={\cal P}\exp\{i\int dz^-\, T^a\, A^+_a({\bf x}_i, z^-)\}
  \label{2.3}
\end{equation}
with $A^+$ - the gauge field in the target and $T^a$ - the
generator of $SU(N_c)$ in a representation corresponding to a given
parton. The relative phases between the components of the wave
function change, and the state that emerges after the target is no
longer an eigenstate of the strong interaction Hamiltonian (as the
incoming state is assumed to be) but rather a superposition
of such eigenstates. It is convenient to rewrite the outgoing wave function in
terms of the second quantized operator corresponding to the gauge
rotation Eq.(\ref{2.3})
\begin{equation}
  \vert\Psi_{out}\rangle=\hat W\vert\Psi_{in}\rangle
  \label{2.4}
\end{equation}
with
\begin{equation}
  \hat W=\exp\left[ i\int \lambda^a({\bf x})\rho^a({\bf x})\right] \, .
  \label{2.5}
\end{equation}
Here $\rho^a({\bf x})$ is the color charge density operator, integrated
over the rapidities of the projectile, and the parameters
$\lambda^a$ are the functions of the target field $A^+$ defined so
that
\begin{equation}
 \exp\left[ iT^a\lambda^a({\bf x}_i)\right] = W({\bf x}_i)\, .
 \label{2.6}
\end{equation}
Various differential cross sections are given by the expectation values of gluonic observables $O(a,a^\dagger)$ in the outgoing wave function at $t\rightarrow\infty$. The evolution between 
the time of scattering and the time of measurement ($t\rightarrow\infty$) is unitary, and thus 
does not affect inclusive observables like total or inelastic cross sections. But it certainly does 
affect the measured particle spectrum via the final state radiation and has to be taken into account. Therefore expectation values of gluonic observables $O(a,a^\dagger)$
are not simply given by  $\langle \Psi_{out}|O(a,a^\dagger)|\Psi_{out}\rangle$ with 
$\vert \Psi_{out} \rangle$ of (\ref{2.4}) since $\vert\Psi_{out}\rangle$ has to be evolved to 
infinite time before the average can be taken~\cite{Kovner:2001vi}. Another way to view the 
final state radiation is to realize that the state immediately after the scattering has a nonvanishing overlap with the incoming state $\vert \Psi_{in} \rangle$
and similar ``hadronic states''.
[By hadronic states in this context we mean the states constructed just like the incoming state, 
so that they are perturbative eigenstates of the QCD Hamiltonian, which in the lowest perturbative 
order are orthogonal to single gluon states. We do not imply of course that within the present 
approximation one has to take into account the nonperturbative hadronization process.] 
This overlap has to be subtracted to get a 
result which counts only the produced free gluons, and not those which are contained in the 
incoming wave function  or the wave function of outgoing ``hadronic'' states
but not freed during the scattering process.

Consider for example the simplest possible projectile, namely
the wave function of a quark at transverse coordinate zero and
color index $\alpha$, which is ``dressed'' to first order in perturbation 
theory. The incoming quark state contains a component coming from the
splitting $\alpha \to \beta\, b$, where the gluon labeled by adjoint index $b$ sits at
a transverse position displaced by ${\bf z}$ from its parent quark. To first
order, the quark wave function reads
\begin{eqnarray}
  \vert\alpha_{D}\rangle&=&\vert \alpha\rangle + \int d{\bf z}\,d\xi f_i({\bf z})\,
  T^b_{\alpha\, \beta}\, \vert \beta\, ; b({\bf z},i,\xi)\rangle
  \nonumber \\
  &=& \vert \alpha\rangle + \int d{\bf z}\, \vec{f}({\bf z})\,
  T^b_{\alpha\, \beta}\, \vert \beta\, ; b({\bf z})\rangle\, .
  \label{2.8}
\end{eqnarray}
Here, the second line specifies a shorthand used below, and 
\begin{equation}
  f_i({\bf z}) = \frac{g}{2\pi^{3/2}} {z_i\over {\bf z}^2}
  \label{2.9}
\end{equation}
denotes the perturbative Weizs\"acker-Williams (WW) gluon field and the index $i$
labels the direction in the transverse plane. For the 
single inclusive gluon cross section, one has to calculate
\begin{equation}
   \langle\delta\Psi\vert {a_i^d}^\dagger({\bf p})\, a_i^d({\bf p})\vert\delta\Psi\rangle\ \ \ ; \ \ \ \ \ \ \ \vert\delta\Psi\rangle
      =\vert \Psi_{out} \rangle
        -\sum_{\alpha}\vert\alpha_{D}\rangle\langle\alpha_{D}\vert\Psi_{out}\rangle\, .
      \label{2.7}
\end{equation}
The subtraction (\ref{2.7}) properly accounts for the evolution of the scattered system after it emerges
from the target~\cite{Kovner:2001vi}, since the only "hadronic state" in the final state to the first order in $\alpha_s$ is the dressed quark. Eq.(\ref{2.7}) ensures that the observable does not include the gluons in the wave function of this dressed quark.

The above example illustrates the general statement that in calculations of 
$O(a,a^\dagger)$, one should not count gluons that belong to the WW 
cloud of any of the fast charged partons that constitute the projectile.
Rather, a "quark" or a "gluon" in the final state should not be thought of as a
free Fock space quark or gluon, but the quark (or gluon) plus its
WW cloud. The change of basis from free to dressed partons is
described by the unitary "gluon cloud" operator $C$ \cite{Kovner:2001vi}
\begin{equation}
   \vert\alpha_{D}\rangle=C\vert\alpha\rangle 
  \label{2.10}
\end{equation}
with
\begin{equation}
  C=P\exp\left(i \int d{\bf x}\, d{\bf z}\,d\xi f_i({\bf z}-{\bf x})\,
    [a_i^d({\bf z},\xi) + a_i^{d\dagger}({\bf z},\xi)]\, 
    \rho^d_\xi({\bf x}) \right)\, ,
    \label{2.11}
\end{equation}
where $P$ stands for rapidity ordering, and $\rho^d_\xi({\bf x})$
denotes the total charge density operator integrated  from the
rapidity of the projectile to $\xi$. The expression (\ref{2.11})
is only valid to first order in $\alpha_s$, but we will not need 
higher orders in the calculations presented in this paper.
The observable
\begin{equation}
C{a_i^d}^\dagger a_i^d C^\dagger
  \label{2.12}
\end{equation}
counts directly those low x gluons whose wave functions are
orthogonal to those of dressed quarks (and dressed "valence"
gluons). Since the dressed gluons are eigenstates of the
propagation outside the target, they can be counted directly in
$\vert \Psi_{out} \rangle$ without the need to account for the additional
unitary evolution to infinite time. This is also true for
any other gluonic observable. Thus the correct expression for calculating 
an arbitrary gluonic observable $O$ in the state $t\rightarrow\infty$ is
\begin{equation}
  \langle \Psi_{out}|CO(a,a^\dagger)C^\dagger|\Psi_{out}\rangle
  \label{2.13}\, ,
\end{equation}
with the gluon cloud operator $C$ given by Eq.(\ref{2.11}).
One can explicitly verify that this procedure reproduces Eq.(\ref{2.7}) in 
the simple example given above.
\section{Single gluon and quark-gluon inclusive cross sections}
\label{sec4}

We now apply the formalism of Section~\ref{sec3} to the calculation of the
one gluon inclusive emission cross section for an arbitrary
projectile which contains a small number of partons, $n\ll
1/\alpha_s$. The restriction to a small number of partons is
necessary to treat the projectile wave function
perturbatively.

We take the wave function of the projectile in the form of a valence Fock space 
dressed by the WW field  
\begin{equation}
     \vert \Psi_{in} \rangle =C\vert R\rangle\, ,
     \label{3.1}
\end{equation}
where $\vert R\rangle$ is an arbitrary state in the free Fock
space with a small number of large rapidity partons. The outgoing
wave function after the propagation through the target is
\begin{equation}
     \vert\Psi_{out}\rangle=\hat WC\vert R\rangle\, .
     \label{3.2}
\end{equation}
The gluon yield per unit rapidity at transverse momentum ${\bf p}$ is given by
[see Eq. (\ref{2.13})]
\begin{equation}
{dN\over dy\, d{\bf p}}=\langle\Psi_{out}\vert
   C{a_i^d}^\dagger({\bf p},y)a_i^d({\bf p},y)C^\dagger\vert\Psi_{out}\rangle\, .
   \label{3.3}
\end{equation}
To evaluate this, we act with the gluon cloud
operator $C$ on the gluon annihilation and creation operators $a$ and $a^\dagger$
\begin{equation}
  Ca_i^dC^\dagger=a_i^d-b_i^d(\rho)\, ,
  \label{3.4}
\end{equation}
where $b_i^d(\rho)$ is the classical WW field associated with the charge
density $\rho$
\begin{equation}
  b^d_i({\bf z},\rho)=\int d{\bf x}
       f_i({\bf z}-{\bf x})\rho^d({\bf x})\, .
       \label{3.5}
\end{equation}
To leading order in $\alpha_s$, the gluon
cloud operator $C$ commutes with the charge density $\rho^d$. We thus
find
\begin{eqnarray}
    &&(a_i-b_i(\rho))\vert\Psi_{out}\rangle=\hat W
          \hat W^\dagger(a_i-b_i(\rho))\hat WC\vert R\rangle\, ,
   \label{3.6}\\
    &&\hat W (wa_i-b_i(w\rho))C\vert R\rangle
      = \hat WC(wa_i-b_i(w\rho))\vert R\rangle\, .
   \label{3.7}
\end{eqnarray}
We have used here the fact that the state $\vert R\rangle$
does not contain gluons with rapidities corresponding to the
operator $a$, and hence it is annihilated by $a$. The shorthand notation in (\ref{3.7})
stands for
\begin{equation} 
  wa_i^d\equiv W_{db}^A({\bf z})a_i^b({\bf z}),\ etc.
  \label{3.8}
\end{equation}
with $W^A$ - the eikonal Wilson factor in the adjoint
representation. We thus have
\begin{eqnarray}
   {dN\over dy\, d{\bf p}}
    &=&\frac{1}{(2\pi)^2} \int d{\bf z} d{\bar {\bf z}}\, 
     e^{-i{\bf p}\cdot ({\bf z}-{\bar {\bf z}})}\langle
     R\vert\left[w^\dagger({\bf z})b_i^d({\bf z},\rho)
                 -b_i^d({\bf z},w\rho)\right]
               \nonumber \\
   && \qquad \qquad  \qquad \qquad \times  
   \left[w({\bar {\bf z}})b_i^d({\bar {\bf z}},\rho)
         -b_i^d({\bar {\bf z}},w\rho)\right]\vert R\rangle\, .
       \label{3.9}
\end{eqnarray}
Using
\begin{eqnarray}
   w({\bf z})b_i^d({\bf z},\rho)-b_i^d({\bf z},w\rho)
   &=& W^A_{db}({\bf z}) \int d{\bf x}f_i({\bf z}-{\bf x})\rho^b({\bf x})
       \nonumber \\
   && - \int d{\bf x}f_i({\bf z}-{\bf x})W^A_{db}({\bf x})\rho^b({\bf x})\, ,
      \label{3.10}
\end{eqnarray}
we find
\begin{eqnarray}
&&{dN\over dy\, d{\bf p}}={\alpha_s C_F\over \pi^2}
\frac{1}{(2\pi)^2} 
\int d{\bf z}d{\bar {\bf z}}
  e^{-i{\bf p}\cdot ({\bf z}-{\bar {\bf z}})}
  \int d{\bf x} d{\bar {\bf x}}
  \langle \rho^a({\bf x})\rho^b({\bar {\bf x}})\rangle_P 
  {({\bf z}-{\bf x})\cdot (\bar {\bf z}-{\bar {\bf x}})
     \over({\bf z}-{\bf x})^2 ({\bar {\bf z}}-{\bar {\bf x}})^2}
  \nonumber\\
&&\langle\left[ {W^A}^\dagger({\bf z})W^A({\bar {\bf z}})
                + {W^A}^\dagger({\bf x}) W^A({\bar {\bf x}})
                - {W^A}^\dagger({\bf z}) W^A({\bar {\bf x}})
                - {W^A}^\dagger({\bf x}) W^A({\bar {\bf z}})\right]^{ab}
          \rangle_T\, ,
  \label{3.11}
\end{eqnarray}
where the averages of the charge density and of the products of
eikonal factors are taken with respect to the projectile and the
target wave functions, respectively. For a translationally invariant and gauge singlet but otherwise arbitrary target 
this expression reduces to the $k_T$-factorized form derived in
\cite{Kovchegov:2001sc,Dumitru:2001ux,Blaizot:2004wu}.

\subsection{Quark-gluon correlation function: $q\, A \to q({\bf k})\, g({\bf p})\, X$}
We now turn to the simplest two-parton correlation function calculable 
in the eikonal formalism: a single quark projectile of final transverse
momentum ${\bf k}$ which shares its recoil between a gluon of transverse
momentum ${\bf p}$ and the target. The correlation function is
\begin{equation}
{dN\over dy\, d{\bf k}\, d{\bf p}}
   = 
   \langle\Psi_{out}\vert
   C{a^b_i}^\dagger({\bf p},y)a^b_i({\bf p},y)
    d^\dagger_\delta({\bf k})d_\delta({\bf k}) C^\dagger\vert\Psi_{out}\rangle_P\, .
   \label{3.20}
\end{equation}
Here, $d^\dagger_\delta$ is the quark creation operator at projectile
rapidity, and $\vert \Psi_{out}(\alpha) \rangle = \hat W C \vert \alpha \rangle$.
We average over the color charge $\alpha$ of the incoming quark,
\begin{equation}
\langle \Psi_{out}\vert O\vert \Psi_{out}\rangle_P
  ={1\over N} \Sigma_\alpha \langle \Psi_{out}(\alpha) \vert
                 O\vert \Psi_{out}(\alpha) \rangle\, .
  \label{3.21}
\end{equation}
The quark color charge density operator is 
\begin{equation}
  \rho^a({\bf x})=d^\dagger_\alpha({\bf x})\, T^a_{\alpha\beta}d_\beta({\bf x})\, .
  \label{3.22}
\end{equation}
This charge creates the Weizs\"acker-Williams field according to Eq.(\ref{3.5}). 
The calculation of the previous subsection is easily 
repeated. The action of the eikonal S-matrix operator $\hat W$ 
amounts to rotating the quark and the gluon creation operators by appropriate eikonal factors. 
This amounts to attaching a fundamental Wilson line 
$W^F_{\beta\delta}({\bf x})$ at the transverse position of the
quark and an adjoint Wilson line $W^A_{bd}({\bf z})$ at the
transverse position of the gluon.
We find
\begin{eqnarray}
 &&a_i^d({\bf z},y)\, d_\delta({\bf x})\, 
   C^\dagger\vert\Psi_{out}(\alpha)\rangle
  \nonumber \\
 && \quad = f_i({\bf z} - {\bf x})
     \left[ T_{\alpha\beta}^b\, W^F_{\beta\delta}({\bf x})
            W^A_{bd}({\bf z}) - T_{\beta\delta}^d\,
            W^F_{\alpha\beta}({\bf x}) \right]\vert\alpha\rangle\, .
          \label{3.23}
\end{eqnarray}
In terms of this expression we can write the quark-gluon correlator (\ref{3.20})
explicitly. In the large-$N$ limit, the color algebra
simplifies considerably and one obtains for the two-particle correlation 
integrated over impact parameter ${\bf b}$
\begin{eqnarray}
\int d{\bf b}\, 
{dN\over dy\, d{\bf k}\, d{\bf p}}
   &=& \frac{1}{(2\pi)^4}  \int_{{\bf x}\, {\bf \bar x}\, {\bf z}\, {\bf \bar z}}
   e^{-i{\bf k}\cdot ({\bf x}-{\bar {\bf x}}) -i{\bf p}\cdot ({\bf z} - {\bar {\bf z}})}
   \vec{f}({\bf z} - {\bf x})\cdot \vec{f}({\bar {\bf z}} - {\bar {\bf x}})
   \nonumber \\
   && \qquad  \times \left[ Q({\bf z},{\bf x},{\bar {\bf x}},{\bar{\bf z}}) 
                    S({\bar {\bf z}},{\bf z}) + S({\bf x},{\bar {\bf x}})
                    \right.
   \nonumber \\
   && \qquad \left. \quad 
                    - S({\bf x},{\bar {\bf z}}) S({\bar {\bf z}},{\bar {\bf x}})
                    - S({\bf x},{\bf z}) S({\bf z},{\bar {\bf x}}) \right]\, .
   \label{3.24}
\end{eqnarray}
Here, we use the shorthand $\int_{\bf x} = \int d{\bf x}$.
This quark-gluon correlation function is expressed 
in terms of two target averages, $S({\bar {\bf z}},{\bf z})$ and
$Q({\bf z},{\bf x},{\bar {\bf x}},{\bar{\bf z}})$ defined in eqs.
(\ref{3.25})-(\ref{5.2}). Integrated over 
${\bf k}$, it reproduces the expression for the single gluon inclusive emission
cross section given in (\ref{3.11}). This can be checked explicitly
by inserting in (\ref{3.11}) the color charge density correlator 
$\langle\rho^a({\bf x})\rho^b({\bar {\bf x}})\rangle_P = {1\over 2N} 
\delta^{ab}\delta({\bf x})\delta({\bar{\bf x}})$ and writing the
averages over adjoint Wilson lines in terms of fundamental ones.

\section{The two gluon inclusive cross section: $n\, A \to g({\bf p}_1)\, g({\bf p}_2)\, X$}
\label{sec5}

In this section we calculate the eikonal cross section for production of
two gluons. We first give the derivation for an arbitrary
perturbative projectile. Then we specialize to the case of a 
single quark projectile.

The cross section for production of two gluons with rapidities 
and transverse momenta $(\eta,{\bf p}_1)$ and $(\xi, {\bf p}_2)$ is
\begin{eqnarray}
{dN\over d\eta d{\bf p}_1d\xi d{\bf p}_2} &=& 
\frac{1}{(2\pi)^4} 
\int_{{\bf z} {\bf \bar z} {\bf u} {\bf \bar u}}
e^{-i{\bf p}_1\cdot ({\bf z}-\bar{\bf z})-i{\bf p}_2\cdot ({\bf u}-{\bar {\bf u}})}
\langle R\vert  
{C^\dagger} {\hat W}^\dagger C
\left[{a_i^a}^\dagger({\bf z},\xi) a_i^a(\bar{\bf z},\xi) \right. 
\nonumber \\
&& \qquad \qquad \qquad \qquad \left. 
 {a_j^b}^\dagger({\bf u},\eta)\, a_j^b(\bar {\bf u},\eta)\right]
 C^\dagger\hat WC\vert R\rangle\, .
 \label{4.1}
\end{eqnarray}
Our convention is such that the rapidity $\eta$ is closer to the
rapidities of the valence partons in the projectile, $\eta\gg\xi$.
We consider $\eta$ and $\xi$ to be sufficiently close, so that no
evolution effects between $\eta$ and $\xi$ have to be taken into
account.

For this case, the action of the cloud operator $C$ Eq.(\ref{2.11}) 
on the gluon field operator can be restricted to the two rapidities 
$\eta$ and $\xi$. 
We can therefore write it schematically as
\begin{equation}
  C=C_\xi C_\eta
  \label{4.2}
\end{equation}
with
\begin {eqnarray}
C_\eta&=&\exp\left[i \int_{ {\bf z}}b^d_i({\bf z})
    [a_i^d({\bf z},\eta) + a_i^{d\dagger}({\bf z},\eta)]\,
    \right]\, ,
    \label{4.3}\\
    C_\xi &=& \exp\left[i\int_{ {\bf z}} 
    [b^d_i({\bf z})+\delta b^d_i({\bf z})]\, [a_i^d({\bf z},\xi) + a_i^{d\dagger}({\bf z},\xi)]\, 
\right]\, .
    \label{4.4}
\end{eqnarray}
Here, $\rho^d({\bf x})$ is the charge density operator at the valence
rapidity. It determines the classical WW field $b_i^d({\bf z})$ as 
specified in Eq. (\ref{3.5}). The color charges produced at rapidity $\eta$
are measured by $\rho_\eta^d({\bf x})$ and constitute an
additional contribution $\delta b^d_i({\bf z})$ to the WW field which affects the 
gluon production at lower rapidities $\xi$, 
\begin{equation}
\delta b^d_i({\bf z})=\int_{\bf x} f_i({\bf z}-{\bf x})\,\rho_\eta^d({\bf x})\, ,
\end{equation}
where
\begin{equation} 
  \rho_\eta^d({\bf x})=a_i^{b\dagger}({\bf x},\eta)
  T^d_{bc}a_i^c({\bf x},\eta)\, , \qquad
  T^a_{bc}=-if^{abc}\, .
  \label{4.5}
\end{equation}
To construct a state with up to two gluons, we have to expand the cloud operator
(\ref{4.2}) up to second order in $\rho$,
\begin{eqnarray}
  C &=& 1+ i \int_{\bf z} \left\{
        [a_i^d({\bf z},\eta)+{a_i^d}^\dagger({\bf z},\eta)]\, 
        b_i^d({\bf z}) \right. 
    \nonumber \\
        && \qquad \qquad 
        + \left. [a_i^d({\bf z},\xi)+{a_i^d}^\dagger({\bf z},\xi)]\, [b_i^d({\bf z})+\delta b_i^d({\bf z})]
                   \right\}\nonumber\\
    && \qquad - \int_{{\bf z},{\bf u}}
       [a_i^d({\bf z},\xi)+{a_i^d}^\dagger({\bf z},\xi)]\,
       [b_i^d({\bf z})+\delta b_i^d({\bf z})]
       \nonumber \\
    && \qquad \qquad \qquad 
       \times [a_i^d({\bf u},\eta)+{a_i^d}^\dagger({\bf u},\eta)]\, b_i^d({\bf u})\, .
          \label{4.6}
\end{eqnarray}
It is now a straightforward albeit tedious matter to calculate
$C^\dagger\hat WC$ and to act with it on $\vert R\rangle$.
[Alternatively, one can construct $\vert \Psi_{\rm in}\rangle$ and
$\vert \Psi_{\rm out}\rangle$ from the cloud operator and evaluate the 
gluon number operator in this $\vert \Psi_{\rm out}\rangle$ state. This
is done in Appendix~\ref{appa}.] Remembering that $\hat W^\dagger a_i^a({\bf z})
\hat W=w^{ab}({\bf z})a_i({\bf z})$ and
$\hat W^\dagger\rho^a({\bf x})W=w^{ab}({\bf x})\rho^b({\bf x})$, we get after some algebra
\begin{eqnarray}
&& {\hat W}^\dagger\, 
   a_i^a({\bar {\bf z}},\xi)\, a_j^b({\bar {\bf u}},\eta)\,  
   C^\dagger\hat WC\vert R\rangle \nonumber \\
&& = \int_{\bar{\bf x}_1,\bar{\bf x}_2} 
     f_i({\bar {\bf z}} - \bar{\bf x}_1)
f_j({\bar {\bf u}} - \bar{\bf x}_2)
     \left\{ \left(w(\bar{\bf x}_1)-w({\bar {\bf z}})\right) \rho(\bar{\bf x}_1)  
     \right\}^a \left\{ w({\bar {\bf u}}) \rho(\bar{\bf x}_2)  
     \right\}^b  \vert R\rangle \nonumber \\
&& - \int_{\bar{\bf x}_1,\bar{\bf x}_2} 
f_i({\bar {\bf z}} - \bar{\bf x}_1)
  f_j({\bar {\bf u}} - \bar{\bf x}_2)
     \left\{ w(\bar{\bf x}_2) \rho(\bar{\bf x}_2) \right\}^b  
     \left\{ \left(w(\bar{\bf x}_1)-w({\bar {\bf z}})\right) \rho(\bar{\bf x}_1)  
     \right\}^a  \vert R\rangle \nonumber \\
&&  + \int_{\bar{\bf x}_1} 
f_i({\bar {\bf z}} - {\bar {\bf u}})
 f_j({\bar {\bf u}} - \bar{\bf x}_1)
     \left\{ \left( w({\bar {\bf z}}) - w({\bar {\bf z}}) \right)
     w^\dagger({\bar {\bf u}})\, T^b\, w({\bar {\bf u}})
     \rho(\bar{\bf x}_1) \right\}^a  \vert R\rangle \, .
     \label{4.7}
\end{eqnarray}
With this state, the expectation value of the observable (\ref{4.1})
can be calculated directly. It contains terms which are quadratic,
cubic and quartic in the density of the projectile,
\begin{equation}
  {dN\over d\eta d{\bf p}_1d\xi d{\bf p}_2}
  = \frac{1}{(2\pi)^4} 
  \int_{{\bf z}\bar{\bf z} {\bf u}\bar{\bf u}} e^{-i{\bf p}_1\cdot ({\bf z}-\bar{\bf z})
                  -i{\bf p}_2\cdot ({\bf u}-\bar{\bf u})}
   \left[\Sigma_2+\Sigma_3+\Sigma_4\right]\, ,
  \label{4.8}
\end{equation}
where
\begin{eqnarray}
\Sigma_2 &=& \int_{{\bf x}\bar{\bf x}}\vec{f}({\bf u} - {\bf z})\cdot \vec{f}({\bar {\bf z}}-{\bar {\bf u}})\, 
\vec{f}({\bar {\bf u}}-\bar{\bf x})\cdot \vec{f}({\bf x} - {\bf u})\,
\langle \rho^a({\bf x})\rho^b(\bar{\bf x})\rangle_P \nonumber \\
&& \times \langle \left\{ w^\dagger({\bf u}) T^c w({\bf u}) 
           \left( w^\dagger({\bf z}) - w^\dagger({\bf u}) \right)
           \right. \nonumber \\
&& \qquad \qquad \left. 
           \left( w({\bar {\bf z}}) - w({\bar {\bf u}}) \right) 
           w^\dagger({\bar {\bf u}}) T^c w({\bar {\bf u}}) \right\}^{ab}  
           \rangle \, ,
           \label{4.9}
\end{eqnarray}
\begin{eqnarray}
\Sigma_3 &=&\int_{{\bf x}_1\bar{\bf x}_1\bar{\bf x}_2} 
             \vec{f}({\bf u} - {\bf z})\cdot \vec{f}({\bar {\bf z}}-\bar {\bf x}_1)\, 
             \vec{f}({\bf x}_1-{\bf u})\cdot \vec{f}({\bar {\bf u}}-\bar{\bf x}_2)
             \nonumber \\
       &&  \times \left[ 
           \langle\rho^c({\bf x}_1)\rho^e(\bar {\bf x}_2)\rho^d(\bar {\bf x}_1)\rangle_P
                      \right.
           \nonumber \\
       && \qquad \qquad \times 
           \langle\left\{[w^\dagger(\bar {\bf x}_1)-w^\dagger({\bar {\bf z}})]
                         [w({\bf z})-w({\bf u})]T^cw^\dagger({\bf u})
                         w(\bar {\bf x}_2)\right\}^{de}\rangle_T
           \nonumber\\
       && \qquad - \langle\rho^c({\bf x}_1)\rho^d(\bar{\bf x}_1)
                           \rho^e(\bar{\bf x}_2)\rangle_P
          \nonumber \\
       && \qquad \qquad \times \left.
          \langle\left\{[w^\dagger(\bar{\bf x}_1)-w^\dagger({\bar {\bf z}})]
           [w({\bf z})-w({\bf u})]T^cw^\dagger({\bf u})
            w({\bar {\bf u}})\right\}^{de}\rangle_T \right]
           \nonumber\\
       &+&\int_{{\bf x}_1{\bf x}_2\bar{\bf x}_1}
           \vec{f}({\bar {\bf z}}-{\bar {\bf u}})\cdot \vec{f}( {\bf x}_1 - {\bf z})\, 
           \vec{f}({\bar {\bf u}}-\bar{\bf x}_1)\cdot \vec{f}({\bf x}_2 - {\bf u})
           \nonumber\\
       &&  \times \left[
           \langle\rho^d({\bf x}_1)\rho^c({\bf x}_2)
                        \rho^e(\bar {\bf x}_1)\rangle_P
                       \right.
           \nonumber \\ 
       && \qquad \qquad \times 
          \langle\left\{ w^\dagger({\bf x}_2) w({\bar {\bf u}}) T^e
           [w^\dagger({\bar {\bf z}})-w^\dagger({\bar {\bf u}})]
           [w({\bf x}_1)-w({\bf z})]\right\}^{cd}\rangle_T
           \nonumber\\
       && \qquad - \langle\rho^c({\bf x}_2)\rho^d({\bf x}_1)
                           \rho^e(\bar{\bf x}_1)\rangle_P
          \nonumber \\
       && \qquad \qquad \times \left.
          \langle\left\{w^\dagger({\bf u}) w({\bar {\bf u}}) T^e
          [w^\dagger({\bar {\bf z}})-w^\dagger({\bar {\bf u}})]
          [w({\bf x}_1)-w({\bf z})]\right\}^{cd}\rangle_T \right] \, ,
      \label{4.10}
\end{eqnarray}
and
\begin{eqnarray}
\Sigma_4 &=& \int_{{\bf x}_1{\bf x}_2\bar{\bf x}_1\bar{\bf x}_2}
\vec{f}({\bar {\bf z}} - \bar{\bf x}_1)\cdot \vec{f}({\bf x}_1-{\bf z})\, 
             \vec{f}({\bar {\bf u}}-\bar{\bf x}_2)\cdot \vec{f}({\bf x}_2-{\bf u})
             \nonumber\\
     && \times \left[
       \langle \rho^g({\bf x}_2) \rho^k({\bf x}_1)
               \rho^l(\bar{\bf x}_1) \rho^h(\bar{\bf x}_2)\rangle_P
               \right. \nonumber \\
     && \qquad \times 
         \langle\left\{[w^\dagger({\bf x}_1)-w^\dagger({\bf z})]
                       [w(\bar{\bf x}_1)-w({\bar {\bf z}})]\right\}^{kl}
                \left\{w^\dagger({\bf u}) w({\bar {\bf u}})\right\}^{gh}
                \rangle_T
                \nonumber\\
     && \quad 
        - \langle \rho^k({\bf x}_1) \rho^g({\bf x}_2)
                  \rho^l(\bar{\bf x}_1) \rho^h(\bar{\bf x}_2)\rangle_P
        \nonumber \\
     && \qquad \times 
         \langle\left\{[w^\dagger({\bf x}_1)-w^\dagger({\bf z})]
                       [w(\bar{\bf x}_1)-w(\bar {\bf z})]
        \right\}^{kl}
        \left\{w^\dagger({\bf x}_2) w({\bar {\bf u}})\right\}^{gh}\rangle_T
        \nonumber\\
     && \quad 
        - \langle\rho^g({\bf x}_2)\rho^k({\bf x}_1)
                 \rho^h(\bar{\bf x}_2)\rho^l(\bar{\bf x}_1)\rangle_P
        \nonumber \\
     && \qquad \times 
        \langle\left\{ [w^\dagger({\bf x}_1)-w^\dagger({\bf z})]
                       [w(\bar{\bf x}_1)-w(\bar {\bf z})]
        \right\}^{kl}
        \left\{w^\dagger({\bf u})w(\bar{\bf x}_2)\right\}^{gh}\rangle_T
        \nonumber\\
     && \quad 
        + \langle\rho^k({\bf x}_1)\rho^g({\bf x}_2)
                 \rho^h(\bar{\bf x}_2)\rho^l(\bar{\bf x}_1)\rangle_P
        \nonumber \\
     && \qquad \times \left. 
        \langle\left\{ [w^\dagger({\bf x}_1)-w^\dagger({\bf z})]
                       [w(\bar{\bf x}_1)-w(\bar {\bf z})]
                    \right\}^{kl}
        \left\{w^\dagger({\bf x}_2)w(\bar{\bf x}_2)\right\}^{gh}\rangle_T 
        \right] \, .
     \label{4.11}
\end{eqnarray}
%
This is an explicit function of the color charge density correlators 
in the projectile, and the correlators of the eikonal factors in the
target. The term $\Sigma_2$ is the probability corresponding to the 
process when the valence component of $\vert R\rangle$ emits the 
gluon with rapidity $\eta$, which subsequently splits into two gluons with
rapidities $\eta$ and $\xi$. The term $\Sigma_4$ corresponds to
the probability of emission of both gluons $\xi$ and $\eta$ from
the valence component, and the term $\Sigma_3$ is the
interference of these two amplitudes. These expressions are general and 
valid for any perturbative projectile. Also note that we have not used the formal 
$1/N_c$ expansion to arrive at these expressions. In these aspects, our expression 
is more general than the one given in \cite{Jalilian-Marian:2004da} for a dipole
as projectile.

\subsection{Two-gluon correlations of a single quark projectile: $q\, A \to g({\bf p}_1)\, g({\bf p}_2)\, X$}
We now specify to the simplest projectile - a single quark. We
take the incoming quark to be at the origin of the transverse
plane. The color charge density operator is (\ref{3.22}) 
and we average again over the color index of the incoming
quark as in (\ref{3.21}).
The color charge density correlators are
\begin{eqnarray}
\langle\rho^a({\bf x})\rho^b({\bf y})\rangle_P &=& {1\over 2N}
                     \delta^{ab}\delta({\bf x})\delta({\bf y}) \, ,
                     \nonumber \\
   \langle\rho^a({\bf x})\rho^b({\bf y})\rho^c({\bf z})\rangle_P &=& 
    {1\over 4N} \left(d^{abc}+if^{abc}\right) 
     \delta({\bf x})\delta({\bf y})\delta({\bf z}) \, ,
     \nonumber\\
  \langle\rho^a({\bf x})\rho^b({\bf y})\rho^c({\bf z})\rho^d({\bf u})\rangle_P &=& 
    \left[   {1\over 2N^2}\delta^{ab}\delta^{cd}
             + {1\over 8N} \left(d^{abe}+if^{abe}\right)
                  \left(d^{ecd}+if^{ecd}\right)
                  \right] \nonumber \\
   && \qquad \times \delta({\bf x}) \delta({\bf y}) \delta({\bf z}) \delta({\bf u})\, .
  \label{4.12}
\end{eqnarray}

The two-gluon correlation function (\ref{4.1}) is given explicitly 
in terms of these projectile-averaged correlators. In the large-$N_c$
limit, the expression simplifies. Using the $SU(N_c)$-identities
compiled in the appendix of \cite{Kovner:2001vi}, the two-gluon correlation
function can be expressed in terms of the target averages 
(\ref{3.25}) and (\ref{3.26}) of two and four fundamental Wilson
lines, respectively.
The final result takes the form
\begin{eqnarray}
 &&{dN\over d\eta d{\bf p}_1d\xi d{\bf p}_2} = 
 \frac{1}{(2\pi)^4} 
 \int_{{\bf z}\bar {\bf z}{\bf u}\bar{\bf  u}}
e^{-i{\bf p}_1\cdot ({\bf z}-\bar {\bf z})
      -i{\bf p}_2\cdot ({\bf u}-\bar {\bf u})}\, \vec{f}({\bf u})\cdot \vec{f}({\bar {\bf u}})
  \nonumber \\
  && \qquad \times \left[ 2\, 
     \vec{f}({\bf z} - {\bf u})\cdot \vec{f}({\bar {\bf z}} - {\bar {\bf u}})
     \left\{ Q({\bar {\bf u}},{\bf u},{\bf z},{\bar {\bf z}})   
             S({\bf z},{\bar {\bf z}})\, 
             S({\bf u},{\bar {\bf u}}) 
             + S^2({\bf u},{\bar {\bf u}}) \right. \right.
     \nonumber \\
  &&  \qquad \qquad \qquad \qquad 
       \left. - S({\bar {\bf u}},{\bar {\bf z}})
              S({\bar {\bf z}},{\bf u}) S({\bf u},{\bar {\bf u}})
              -S({\bf u},{\bf z}) S({\bf z},{\bar {\bf u}})
              S({\bar {\bf u}},{\bf u})
              \right\} 
     \nonumber \\
  &&  \qquad \quad -  \vec{f}({\bf z} - {\bf u})\cdot \vec{f}({\bar {\bf z}})
     \left\{ Q({\bf u},{\bar {\bf u}},{\bar {\bf z}}, {\bf z})   
             S({\bf u},{\bar {\bf u}}) S({\bf z},{\bar {\bf z}}) 
             + S({\bf u}) S({\bar {\bf u}}) 
               S({\bf u},{\bar {\bf u}}) \right. 
     \nonumber \\
  &&  \qquad \qquad \qquad \qquad 
             - Q({\bf u},{\bar {\bf u}},{\bf 0}, {\bf z})   
             S({\bf z})\, S({\bf u},{\bar {\bf u}}) 
             - S({\bar {\bf u}},{\bf u}) 
               S({\bf u},{\bar {\bf z}}) 
               S({\bar {\bf z}},{\bar {\bf u}}) 
     \nonumber \\
  &&  \qquad \qquad \qquad \qquad 
             + Q({\bf u},{\bf 0},{\bar {\bf z}},{\bf z})   
             S({\bf z},{\bar {\bf z}})\, S({\bf u}) 
             + S^2({\bf u})
     \nonumber \\
  &&  \qquad \qquad \qquad \qquad \left. 
             - S({\bf u}) S({\bf z}) S({\bf z},{\bf u})
             - S({\bf u}) S({\bar {\bf z}}, {\bf u}) S({\bar {\bf z}})
             \right\}
    \nonumber \\
  &&  \qquad \quad -  \vec{f}({\bf z})\cdot \vec{f}({\bar {\bf z}}- {\bar {\bf u}})
     \left\{ Q({\bar {\bf u}},{\bf u}, {\bf z},{\bar {\bf z}})   
             S({\bf u},{\bar {\bf u}}) S({\bf z},{\bar {\bf z}}) 
             + S({\bf u}) S({\bar {\bf u}}) 
               S({\bf u},{\bar {\bf u}}) \right. 
     \nonumber \\
  &&  \qquad \qquad \qquad \qquad 
             - Q({\bar {\bf u}},{\bf u},{\bf 0}, {\bar {\bf z}})   
             S({\bar {\bf z}})\, S({\bf u},{\bar {\bf u}}) 
             - S({\bar {\bf u}},{\bf u}) 
               S({\bf z},{\bar {\bf u}}) 
               S({\bf z},{\bf u}) 
     \nonumber \\
  &&  \qquad \qquad \qquad \qquad 
             + Q({\bar {\bf u}},{\bf 0},{\bf z},{\bar {\bf z}})   
             S({\bf z},{\bar {\bf z}})\, S({\bar {\bf u}}) 
             + S^2({\bar {\bf u}})
     \nonumber \\
  &&  \qquad \qquad \qquad \qquad \left. 
             - S({\bar {\bf u}}) S({\bar {\bf z}}) 
               S({\bar {\bf z}},{\bar {\bf u}})
             - S({\bar {\bf u}}) S({\bf z}) S({\bf z},{\bar {\bf u}})
             \right\}
     \nonumber \\
  &&  \qquad \quad +  \vec{f}({\bf z})\cdot \vec{f}({\bar {\bf z}})
      \left\{ Q({\bar {\bf u}},{\bf u},{\bf z},{\bar {\bf z}})   
              S({\bf z},{\bar {\bf z}})\, 
              S({\bf u},{\bar {\bf u}}) 
              + S^2({\bf u},{\bar {\bf u}})\right.
     \nonumber \\
  &&  \qquad \qquad \qquad \qquad 
             - Q({\bar {\bf u}},{\bf u},{\bf 0},{\bar {\bf z}})   
             S({\bf u},{\bar {\bf u}})\, S({\bar {\bf z}}) 
             + S^2({\bf z},{\bar {\bf z}})
     \nonumber \\
  &&  \qquad \qquad \qquad \qquad \left. \left.
             - Q({\bar {\bf u}},{\bf u},{\bf z},{\bf 0})   
             S({\bf u},{\bar {\bf u}})\, S({\bf z}) +1 
             - S^2({\bar {\bf z}}) - S^2({\bf z}) \right\} \right] \, .
         \label{4.13}
\end{eqnarray}
We note that the two gluon correlation function of the perturbative $q \bar{q}$ dipole 
projectile was derived in Ref.~\cite{Jalilian-Marian:2004da} in the large $N_c$ limit.
 We have 
checked that in the limit of arbitrary large dipole size,
the final result of Ref. {\cite{Jalilian-Marian:2004da} coincides with (\ref{4.13}). 

\section{Conclusions}

In this paper, we have shown that the framework of perturbative saturation
 provides for a mechanism which can shift the maximum strength which of azimuthal particle correlations 
 away from $180^\circ$.
We understand the physics of this phenomenon in the following way. 
There are two basic mechanisms by which a quark and a gluon can be produced 
in the final state. The first one is the QCD analog of Moli\`ere scattering. 
The quark in the projectile wave function
scatters from the perturbative part of the target gluon field and then radiates 
a gluon in the final state. This final state radiation is predominantly back-to-back
and generates maximal correlations at angle $\pi$. The second mechanism is the 
multiple scattering of the $qg$ component of the initial state on the target. 
The final states produced by this mechanism depend strongly on the separation 
between the initial quark and gluon in the transverse plane. For components 
with large separation (small 
relative transverse momentum $|{\bf k}-{\bf p}|\ll Q_s$ 
{\it in the initial state}), the quark and the gluon scatter independently off 
the target fields. This process produces uncorrelated
$qg$ pairs in the final state and leads to the broadening of the angular distribution. 
However, the components of the incoming state which have transverse size smaller than 
the correlation length of the target fields scatter coherently, i.e. they scatter
effectively as a single particle. These quark-gluon components pick up a typical
soft momentum (of order $Q_{s,0}$, as explained in Section~\ref{sec3}), which is equally 
shared between the two partons. As a result, the two partons emerge from the interaction 
region with momenta shifted in the same direction by an equal amount of order 
$Q_{s,0}/2$. In the initial state, the total transverse momentum vanishes, and thus 
the momenta of the quark and the gluon are balanced. The equal momentum transfer 
therefore produces angular correlation in the final state which is peaked 
away from $180^\circ$. The maximum correlation produced by this mechanism 
for large trigger momenta $k$ can be simply estimated to lie about $Q_{s,0}/\sqrt{2}k$ 
away from $180^\circ$. Whether this shift in the maximal correlation angle is actually 
observable in the spectrum depends very much on the relative importance of the 
Moli\`ere component and the coherent soft scattering component. Our numerical 
results suggest that the coherent scattering component is dominant for the trigger 
momenta around $Q_s$. 

We also observe that the coherent scattering component, which shifts the maximal
correlation strength away from $180^\circ$, is enhanced and dominates over a wider
range of trigger momenta, if the projectile system has smaller transverse size. 
To establish this statement, we use the fact that our calculation allows us to
regulate the transverse size of the incoming projectile via the cut-off $\Lambda_{\rm cut}$.
Although it is difficult to draw firm phenomenological conclusions, we believe that there
is potential for an observable effect.

Also, it would be interesting to study how this picture discussed above is affected by 
low-$x$ evolution. One may expect that since $Q_s$ increases with rapidity, the
maximal correlation angle at forward rapidities should move further away from $180^\circ$. 
However, low-$x$ evolution does not only affect the value of the saturation
scale, but also changes significantly the efficiency of the target as a function of 
momentum transfer. In particular, the momentum dependence of the target gluon distribution decreases slower at large momenta thus effectively increasing the Moli\`ere hard scattering component. Hence, it is not clear to us how the azimuthal shift of the maximal
two-parton correlation strength will evolve with $x$.

\appendix
\section{Alternative calculation of two-gluon correlation function (\ref{4.13})}
\label{appa}

In this appendix, we derive the two-gluon correlation function (\ref{4.13}) 
following the formulation given in ~\cite{Kovner:2001vi,Kovner:2003zj}.
We start from the state $\vert \psi_{\rm in}^\alpha \rangle$ in (\ref{2.1}) of a 
single quark with color $\alpha$ and its gluon cloud, expanded in perturbation 
theory up to $O(g^2)$,
%
\begin{figure}[h]
\begin{center}
\includegraphics[width=10.5cm]{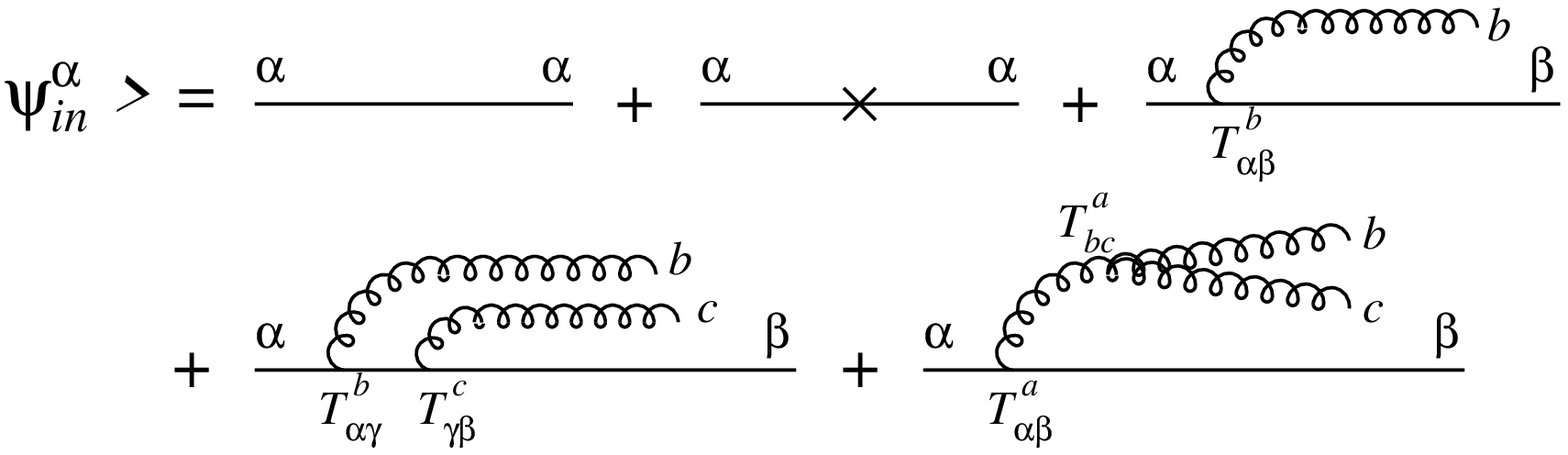}
\end{center}
\end{figure}
%
%
This is the incoming state with two gluons of color $b$ and $c$. 
The crossed line indicates a probability conserving virtual correction,
such that the state $\vert \psi_{\rm in}^{\alpha}\rangle$ is normalized
to unity up to $O(g^2)$. In this approximation, the states do not depend 
on rapidity, and the rapidity labels are suppressed in the following.

The interaction of $\vert \psi_{\rm in}^{\alpha}\rangle$ with the target
results in phase shifts by eikonal Wilson lines, which are in the fundamental
($W^F$) and adjoint ($W^A$) representation, respectively. This leads to the
outgoing state
\begin{eqnarray}
        \vert \Psi_{\rm out}^\alpha \rangle  &=&
        \left( 1 - \frac{C_F}{2} \int d{\bf x}\, \vec{f}({\bf x})\cdot \vec{f}({\bf x})  \right)\, 
         W_{\alpha\beta}^F({\bf 0})\, \vert \beta \rangle
         \nonumber \\    
         && + i \int d{\bf x}\,  \vec{f}({\bf x})
                \left( T^a\, W^F({\bf 0}) \right)_{\alpha\beta}\, W_{ab}^A({\bf x})\,
                \vert \beta; b({\bf x})\rangle   
         \nonumber \\
         && - \frac{1}{2} \int d{\bf x}\, d{\bf y}\, \left[  \vec{f}({\bf x}) \vec{f}({\bf y})
                \left( T^a\, T^d\, W^F({\bf 0}) \right)_{\alpha\beta}
                W_{ab}^A({\bf x})\, W_{dc}^A({\bf y}) \right.
                \nonumber \\
                && + \left. \vec{f}({\bf x}) \vec{f}({\bf y}-{\bf x})\, 
                        \left(T^e\, W^F({\bf 0})\right)_{\alpha\beta}
                        \left( T_{ad}^e W_{ab}^A({\bf x})\, W_{dc}^A({\bf y})\right) \right]
                        \vert \beta; b({\bf x}), c({\bf y})\rangle\, .
\end{eqnarray}
The quark propagates at the transverse position ${\bf x}_q=0$, and the gluons
propagate at ${\bf x}$ and ${\bf y}$, respectively. 

According to (\ref{2.7}), we have to subtract the overlap with the dressed incoming
state. This leads to the state $\vert \delta \Psi_{\alpha} \rangle$ which to lowest order
in $\vec{f} \vec{f}$ consists of the following contributions:
%
\begin{figure}[h]
\begin{center}
\includegraphics[width=8.5cm]{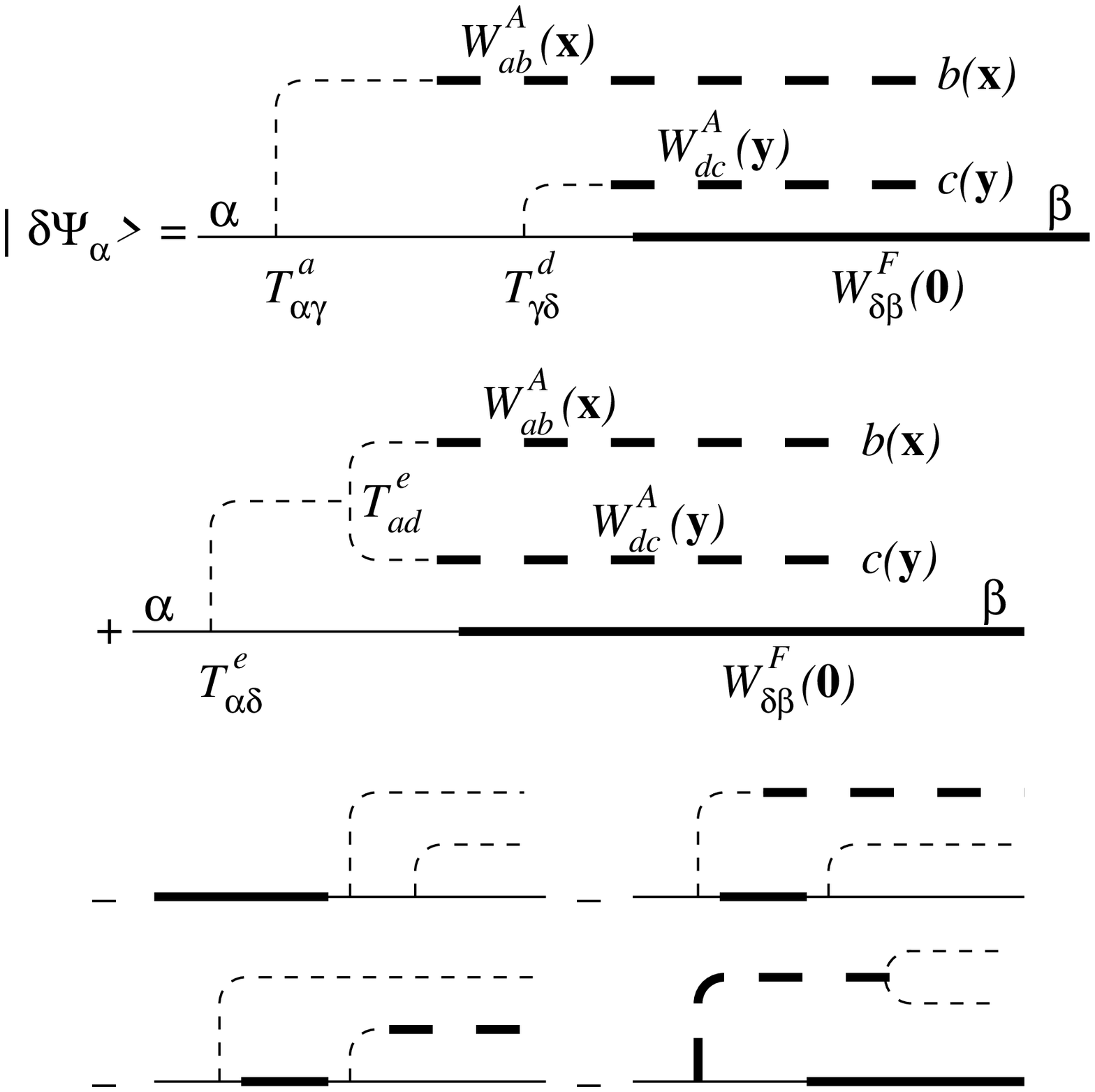}
\end{center}
\end{figure}
%
Here, we use the diagrammatic shorthand of Ref.~\cite{Kovner:2001vi} in which
thick lines denote partons propagating through the target and thus picking up
eikonal Wilson lines in the corresponding representation. The first two graphs 
correspond to the case that the quark emits both gluons before the target and 
all three partons propagate through the target (thick lines). The third diagram
can be viewed as emission after interaction since the outgoing lines do not
carry eikonal factors (thin lines). This third diagram, as well as the remaining
three arise from the subtraction of the overlap in (\ref{2.7}). 

A convenient expression for $\vert \delta \Psi_{\alpha} \rangle$ is obtained by
applying identities like 
\begin{equation}
\left( T^b\, W^F({\bf 0})\, T^d\right)_{\alpha\beta}
= \left( T^b\, T^a\, W^F({\bf 0}) \right)_{\alpha\beta} W_{ad}^A({\bf 0})\, .
\end{equation}
This allows us to shift the Wilson operator $W^F({\bf 0})$ such that
we can write 
\begin{eqnarray}
        \vert \delta \Psi_{\alpha} \rangle &=& - \int d{\bf x}\, d{\bf y}\, \vec{f}({\bf x})\, \vec{f}({\bf y})
                \left( T^a\, T^d\, W^F({\bf 0}) \right)_{\alpha\beta}
                \left[ W_{ab}^A({\bf x}) \left( W_{dc}^A({\bf 0}) - W_{dc}^A({\bf y})   \right) 
                \right. \nonumber\\
                && \qquad \left.
                - W_{db}^A({\bf 0}) \left( W_{ac}^A({\bf 0}) - W_{ac}^A({\bf y})   \right) \right]
                \vert \beta; b({\bf x}), c({\bf y}) \rangle
                \nonumber \\
                && + \int d{\bf x}\, d{\bf y}\, \vec{f}({\bf x})\, \vec{f}({\bf y}-{\bf x})\, 
                         \left(T^e\, W^F({\bf 0}) \right)_{\alpha\beta}
                         \left[T_{ad}^e W_{ab}^A({\bf x})  W_{dc}^A({\bf y})
                 \right. \nonumber\\
                 && \qquad \left.  - W_{ea}^A({\bf x}) T_{bc}^a \right]\, 
                \vert \beta; b({\bf x}), c({\bf y}) \rangle\, .
\end{eqnarray}
This expression may be written in the shorthand form, introducing an operator
${\cal O}({\bf x},{\bf y})$ as follows
\begin{equation}
        \vert \delta \Psi_{\alpha} \rangle = \int d{\bf x}\, d{\bf y}\, 
        {\cal O}_{\alpha\beta}^{bc}({\bf x},{\bf y})\,
                \vert \beta; b({\bf x}), c({\bf y})\rangle\, .
\end{equation}
This allows us to write the two-gluon correlation in a compact way by calculating the expectation
value of the two-particle number operator in this state, averaged over the incoming quark color 
index $\alpha$,
\begin{eqnarray}
        && \frac{1}{N} \sum_{\alpha} \langle \delta \Psi_{\alpha} \vert 
          a^\dagger({\bf p}_1,\xi) a({\bf p}_1,\xi)  a^\dagger({\bf p}_2,\eta) a({\bf p}_2,\eta) 
           \vert \delta \Psi_{\alpha} \rangle 
           \nonumber \\
           && =
           \frac{1}{(2\pi)^4} \int_{{\bf z}\bar{\bf z}{\bf u}\bar{\bf u} } 
           e^{-i{\bf p}_1\cdot ({\bf z}-\bar{\bf z}) - i{\bf p}_2\cdot ({\bf u}-\bar{\bf u})}
           \langle  \left[ {\cal O}_{\alpha\beta}^{bc}(\bar{\bf z},\bar{\bf u}) \right]^\dagger 
           {\cal O}_{\alpha\beta}^{bc}({\bf z},{\bf u}) 
           \rangle_T\, .
\end{eqnarray}
Working out the target averages as described in \cite{Kovner:2001vi}, we obtain after
some color algebra the result given in (\ref{4.13}).



\begin{thebibliography}{99}
%
\bibitem{Arsene:2004fa}
  I.~Arsene {\it et al.}  [BRAHMS Collaboration],
  arXiv:nucl-ex/0410020.
%
\bibitem{Back:2004je}
  B.~B.~Back {\it et al.} [PHOBOS Collaboration],
  arXiv:nucl-ex/0410022.
%
\bibitem{Adcox:2004mh}
  K.~Adcox {\it et al.}  [PHENIX Collaboration],
  arXiv:nucl-ex/0410003.
%
\bibitem{Adams:2005dq}
  J.~Adams {\it et al.}  [STAR Collaboration],
  arXiv:nucl-ex/0501009.
  %
\bibitem{Mueller:1999yb}
  A.~H.~Mueller,
 {\it Lectures given at NATO Advanced Study Institute on Particle Production Spanning MeV and TeV Energies (Nijmegen 99), Nijmegen, Netherlands, 8-20 Aug 1999},
  arXiv:hep-ph/9911289.
 %
\bibitem{Blaizot:2004px}
  J.~P.~Blaizot and F.~Gelis,
  Nucl.\ Phys.\ A {\bf 750} (2005) 148.
%
\bibitem{McLerran:2005kk}
  L.~McLerran,
  Nucl.\ Phys.\ A {\bf 752}, 355 (2005).
 %
\bibitem{Iancu:2003xm}
  E.~Iancu and R.~Venugopalan,
  in ``Quark Gluon Plasma 3'', (editors: R.C. Hwa and X.N. Wang,
   World Scientific, Singapore), p.249-363, 
  arXiv:hep-ph/0303204.
  %
\bibitem{Weigert:2005us}
  H.~Weigert,
  arXiv:hep-ph/0501087.
%
\bibitem{Jalilian-Marian:2005jf}
  J.~Jalilian-Marian and Y.~V.~Kovchegov,
  arXiv:hep-ph/0505052.
 %
\bibitem{Jacobs:2004qv}
  P.~Jacobs and X.~N.~Wang,
  Prog.\ Part.\ Nucl.\ Phys.\  {\bf 54}, 443 (2005).
%
\bibitem{Baier:2000mf}
  R.~Baier, D.~Schiff and B.~G.~Zakharov,
  Ann.\ Rev.\ Nucl.\ Part.\ Sci.\  {\bf 50}, 37 (2000).
%
\bibitem{Kovner:2003zj}
  A.~Kovner and U.~A.~Wiedemann,
  in ``Quark Gluon Plasma 3'', (editors: R.C. Hwa and X.N. Wang,
   World Scientific, Singapore), P.192-248,
  arXiv:hep-ph/0304151.
%
\bibitem{Gyulassy:2003mc}
  M.~Gyulassy, I.~Vitev, X.~N.~Wang and B.~W.~Zhang,
  in ``Quark Gluon Plasma 3'', (editors: R.C. Hwa and X.N. Wang,
   World Scientific, Singapore), p. 123-191,
  arXiv:nucl-th/0302077.
%
\bibitem{Salgado:2003rv}
  C.~A.~Salgado and U.~A.~Wiedemann,
  Phys.\ Rev.\ Lett.\  {\bf 93}, 042301 (2004).
 %
\bibitem{Armesto:2004pt}
  N.~Armesto, C.~A.~Salgado and U.~A.~Wiedemann,
  Phys.\ Rev.\ Lett.\  {\bf 93}, 242301 (2004).
 %
\bibitem{Casalderrey-Solana:2004qm}
  J.~Casalderrey-Solana, E.~V.~Shuryak and D.~Teaney,
  arXiv:hep-ph/0411315.
 %
\bibitem{Shuryak:2004cy}
  E.~V.~Shuryak,
  Nucl.\ Phys.\ A {\bf 750} (2005) 64.
%
\bibitem{Kharzeev:2004bw}
  D.~Kharzeev, E.~Levin and L.~McLerran,
  Nucl.\ Phys.\ A {\bf 748} (2005) 627.
%
\bibitem{Kovner:2001vi}
A.~Kovner and U.~A.~Wiedemann,
Phys.\ Rev.\ D {\bf 64} (2001) 114002.
%
\bibitem{Jalilian-Marian:2004da}
  J.~Jalilian-Marian and Y.~V.~Kovchegov,
  Phys.\ Rev.\ D {\bf 70}, 114017 (2004)
  [Erratum-ibid.\ D {\bf 71}, 079901 (2005)].
%
\bibitem{Nikolaev:2004cu}
  N.~N.~Nikolaev and W.~Sch\"afer,
  Phys.\ Rev.\ D {\bf 71} (2005) 014023.
 %
\bibitem{Nikolaev:2005dd}
  N.~N.~Nikolaev, W.~Sch\"afer, B.~G.~Zakharov and V.~R.~Zoller,
  arXiv:hep-ph/0504057.
  %
\bibitem{Fujii:2005vj}
  H.~Fujii, F.~Gelis and R.~Venugopalan,
  arXiv:hep-ph/0504047.
 %
\bibitem{Baier:2003hr}
R.~Baier, A.~Kovner and U.~A.~Wiedemann,
Phys.\ Rev.\ D {\bf 68}, 054009 (2003).
%
\bibitem{Kharzeev:2002pc}
D.~Kharzeev, E.~Levin and L.~McLerran,
Phys.\ Lett.\ B {\bf 561}, 93 (2003).
%
\bibitem{Albacete:2003iq}
J.~L.~Albacete, N.~Armesto, A.~Kovner, C.~A.~Salgado and U.~A.~Wiedemann,
Phys.\ Rev.\ Lett.\  {\bf 92}, 082001 (2004).
%
\bibitem{Jalilian-Marian:2003mf}
J.~Jalilian-Marian, Y.~Nara and R.~Venugopalan,
Phys.\ Lett.\ B {\bf 577}, 54 (2003).
%
\bibitem{Kharzeev:2003wz}
D.~Kharzeev, Y.~V.~Kovchegov and K.~Tuchin,
Phys.\ Rev.\ D {\bf 68}, 094013 (2003).
%
\bibitem{Kovchegov:1998bi}
Y.~V.~Kovchegov and A.~H.~Mueller,
Nucl.\ Phys.\ B {\bf 529} (1998) 451.
%
\bibitem{Bethe}
 H.~A.~Bethe, Phys.\ Rev.\ D{\bf 89} (1953) 1256.  
%
\bibitem{Kovchegov:2001sc}
Y.~V.~Kovchegov and K.~Tuchin,
Phys.\ Rev.\ D {\bf 65}, 074026 (2002).
%
\bibitem{Dumitru:2001ux}
  A.~Dumitru and L.~D.~McLerran,
  Nucl.\ Phys.\ A {\bf 700}, 492 (2002).
%
\bibitem{Blaizot:2004wu}
J.~P.~Blaizot, F.~Gelis and R.~Venugopalan,
Nucl.\ Phys.\ A {\bf 743}, 13 (2004).
\end{thebibliography}
\end{document}